\documentclass
[pra,superscriptaddress,showpacs, twocolumn, aps,floatfix]{revtex4-1}%
\usepackage{hyperref}
\usepackage[pdftex]{graphicx}
\usepackage{bm}
\usepackage{etoolbox}
\usepackage[usenames,dvipsnames]{color}
\usepackage{amsmath}
\usepackage{amsfonts}
\usepackage{amssymb}%
\usepackage{graphicx}
\usepackage{dcolumn}
\usepackage{bm}
\usepackage{verbatim}
\usepackage{color}
\usepackage{ulem}
\usepackage{epsf}
\usepackage{graphicx}
\newcommand{\bg}{ \begin{gather} }
\newcommand{\eg}{\end{gather}}
\newcommand{\be}{ \begin{equation} }
\newcommand{\ee}{\end{equation}}
\newcommand{\bea}{ \begin{eqnarray} }
\newcommand{\eea}{\end{eqnarray}}

\def \m {\mathbf }

\def \be {\begin{equation}}
\def \ee {\end{equation}}
\def \bea {\begin{align}}
\def \eea {\end{align}}

\def \BEA {\begin{eqnarray}}
\def \EEA {\end{eqnarray}}

\def \BC {\begin{cases}}
\def \EC {\end{cases}}

\begin{document}
\title{
Hydrodynamic Inverse Faraday Effect in Two Dimensional Electron Liquid
}

\author{S.\,O.~Potashin}
\affiliation{Ioffe  Institute, 194021 St.~Petersburg, Russia}

\author{ V. Yu.~Kachorovskii }
\affiliation{Ioffe  Institute, 194021 St.~Petersburg, Russia} \affiliation{ Rensselaer Polytechnic Institute,12180, Troy, NY, USA} \affiliation{CENTERA Laboratories, Institute of High Pressure Physics, Polish Academy of Sciences, 01-142 Warsaw, Poland }
\author{M. S.~Shur}
\affiliation{ Rensselaer Polytechnic Institute,12180, Troy, NY, USA}

\begin{abstract}

We show that a small conducting object, such as a nanosphere or a nanoring, embedded into
or placed in the vicinity of the two-dimensional electron liquid (2DEL)
and subjected to a circularly polarized electromagnetic radiation
induces ``twisted'' plasmonic oscillations in the adjacent 2DEL. The
oscillations are rectified due to the hydrodynamic nonlinearities
leading to the helicity sensitive circular dc current and to a magnetic
moment. This hydrodynamic inverse Faraday effect (HIFE) can be observed
at room temperature in different  materials. The
HIFE is dramatically enhanced in a periodic array of the nanospheres
forming a resonant plasmonic coupler. Such a coupler exposed to a
circularly polarized wave converts the entire 2DEL into a vortex state.
Hence, the twisted plasmonic modes support resonant plasmonic-enhanced
gate-tunable optical magnetization. Due to the interference of the
plasmonic and Drude contributions, the resonances have an asymmetric
Fano-like shape. These resonances present a signature of the 2DEL
properties not affected by contacts and  interconnects and, therefore,
providing the most accurate information about the 2DEL properties. In
particular, the widths of the  resonances encode direct information
about  the momentum relaxation time and viscosity of the 2DEL.

\end{abstract}
\maketitle
\section{Introduction}
\label{s1}

Generation of stationary magnetic moment by a circularly polarized radiation is commonly
referred to as the inverse Faraday effect (IFE) predicted by Pitaevskii
\cite{Pitaevskii61} and first observed by van der Ziel et al. \cite{Ziel65}. Although this effect is
usually studied in magnetic materials \cite{Kimel05,Kirilyuk10,Kirilyuk11}, it can be also observed in
conventional semiconductor nanostructures such as quantum dots and
nanorings \cite{Kibis11,Kibis13,Alexeev13,Joibari14,Alexeev12,Kruglyak2005,Kruglyak2007,Polianski2009,Koshelev15,
Koshelev17}. In particular,   it was  recently predicted \cite{Koshelev15,
Koshelev17}   that a circularly polarized
radiation  with  the electric component $\mathbf  E = \mathbf E_\omega
\exp(-i\omega t )+ c.c.$
 can excite a circular dc current in a nanoring, which, in turn, generates
a magnetic moment
\be \boldsymbol{M}
\propto i\;
\mathbf E _\omega \times  \mathbf  E^*_\omega. \label{M} \ee
 The  proportionality coefficient in Eq.~\eqref{M}  is an odd function of frequency,
  so that the effect is sensitive to the helicity of polarization.
Remarkably, IFE is dramatically
enhanced in vicinity of plasmonic resonances \cite{Koshelev17}. Specifically,
adjusting the plasmonic frequency in the nanoring to match the frequency
of impinging radiation results in much larger optically-induced stationary magnetic
field (up to 0.1 Gauss for typical parameters of a nanoring, see discussion in Ref.~\cite{Koshelev17}).
Hence, an array  of nearly identical quantum rings should give rise to large
  optically-controlled macroscopic magnetization.
This opens a wide
avenue for applications in tunable  optoelectronics, in particular,
in the terahertz (THz) range of frequencies.

The key feature  of  the plasmonic-enhanced  IFE as compared to other plasma wave related
effects is the   absence of the symmetry limitations for conversion of incoming radiation into a dc signal.
Indeed,    in conventional plasmonic devices  such conversion requires
an asymmetry of the system that determines
direction  of the dc current. In the two-dimensional structures,
 the asymmetry can be created by the boundary conditions \cite{Dyakonov93} or
induced by  ratchet effect (see Ref.~\onlinecite{Ivchenko2011} for review). The latter implies a
special type of grating-gate couplers that could provide the required
asymmetry.  By contrast,  IFE exists in fully symmetric rings \cite{Koshelev15,Koshelev17},
and direction of the arising dc current is simply determined by the sign of the circular  polarization.
What is also important in view of  possible  applications for the THz plasmonics,
the  optically-induced dc current remains finite even in the longwavelength limit, when
  $\m E_\omega$  does not vary within the dimension of  ring.
  Hence, the quantum nanorings and ring-based arrays can be used as an effective
  helicity-driven sensors for THz radiation [see estimates and discussion in Ref.~\cite{Koshelev17}].
%
\begin{figure}[b]
\centerline{\includegraphics[width=0.4\textwidth]{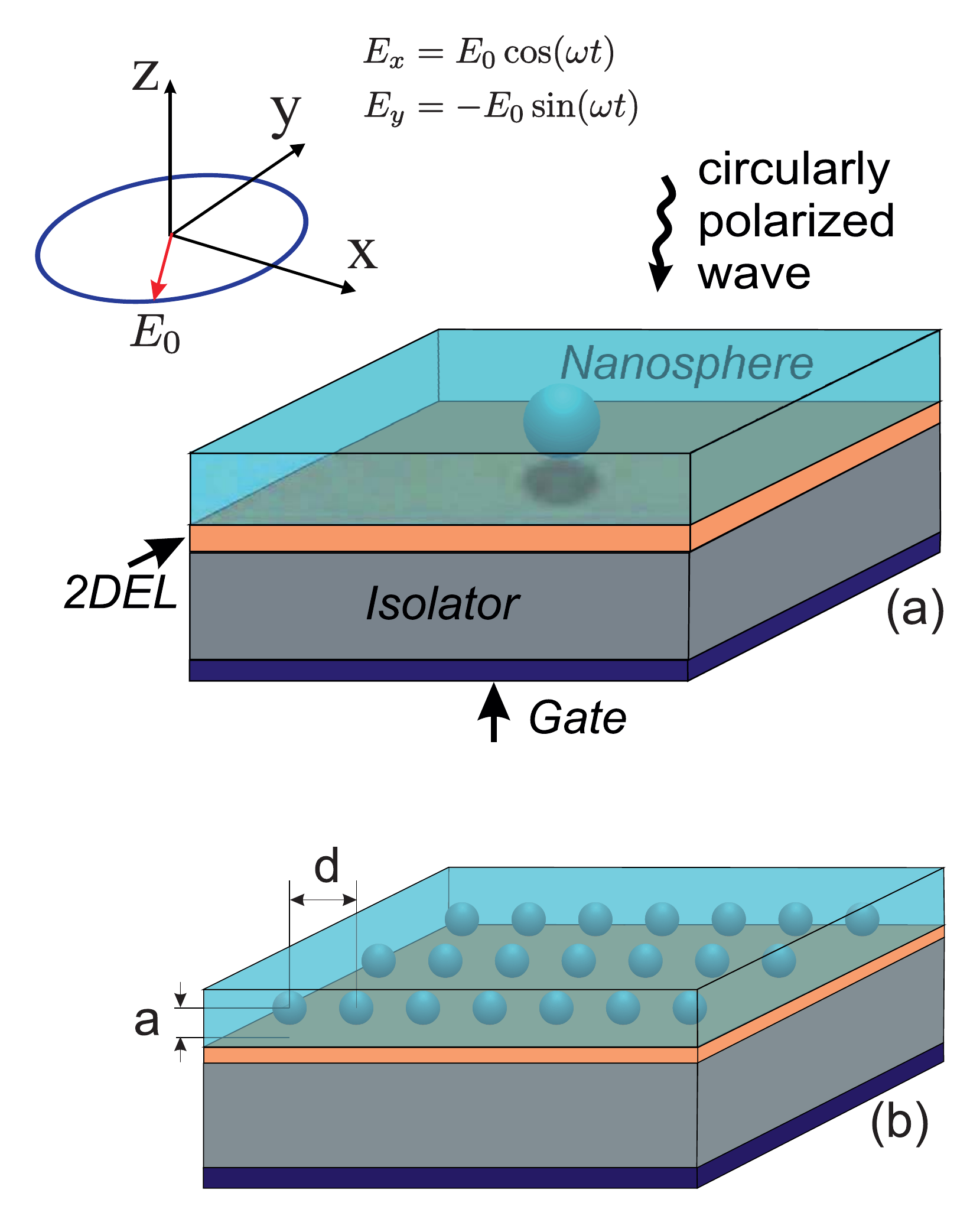}}
\caption{Excitation of twisted plasmons in 2D electron liquid by a single nanosphere
embedded into dielectric matrix  and excited by circularly polarized radiation (a) or
by an array of nanospheres forming plasmonic coupler (b)   } \label{Fig1}
\end{figure}

In this paper, we discuss the possibility of observing similar effects in 2D systems.
We consider the excitation of circular plasmonic modes
(``twisted plasmons'') and circular dc currents in two dimensional electronic
liquid (2DEL). These modes are excited by a circularly polarized
electromagnetic radiation impinging on the metallic or semiconducting
nanosphere or nanoring embedded into or placed above the 2DEL and
inducing rotating dipoles in these nanostructures (see Fig.~\ref{Fig1}a).
    Rectification of the twisted plasmons due to hydrodynamic
nonlinearities leads to a helicity-sensitive circular DC current, and
consequently, to a magnetic moment, thus demonstrating the hydrodynamic
inverse Faraday effect (HIFE). If the nanospheres form a 2D crystal (see Fig.~\ref{Fig1}b), only
the plasmons with the wave vectors forming inverse crystal lattice are
excited, so that excitation spectrum becomes discrete. When
 the radiation frequency is close to any of the discrete plasmonic
frequencies, the entire high-mobility 2DEL experiences a resonant circular
plasmonic excitation. The rectification of these oscillations leads to
  plasmonic-enhanced   DC current which
oscillates in space. The circular dc current and magnetic moment
generated by this current show sharp HIFE resonances. Since the plasma
wave frequency is tunable by the gate voltage and by an external magnetic
field such a system can be used for the optical tunable magnetization of 2D
systems. The typical 2DEL twisted plasmon frequencies are in the THz range,
 and this coupling system could be used for tunable THz
electronic components, including frequency multipliers, modulators,
absorbers, and mixers. Another key application is in the contactless
characterization and parameter extraction of the 2DEL.

Apart from these applications, there are some very interesting  fundamental aspects
of the HIFE related to hydrodynamic approach in plasmonics, the field
which explores how electromagnetic fields can be confined
over dimensions much  smaller than the  radiation wavelength
\cite{Rev1,Rev2,Rev3,Rev4,Rev5,Rev6,Rev7}.  The hydrodynamic approach to the description
of the electronic systems and, in particular, the plasma wave excitation,
has a long history which can be
traced back to the early work by Gurzhi \cite{Gurzhi} and by Jong and Molenkamp
\cite{Molenkamp}, where  hydrodynamic effects on the electron and phonon transport were
discussed and to the work by Dyakonov and Shur \cite{Dyakonov93}, which exploited
the analogy between the  ``shallow water''  hydrodynamics and
that of the electron liquid in the  two-dimensional (2D) gated systems.  Many  other
beautiful hydrodynamic phenomena such as choking of electron flow
\cite{2}, nonlinear rectification of the plasma waves \cite{3,4} and the
formation of the plasmonic shock waves \cite{5} have been subsequently
proposed. Possible applications of these phenomena to the plasma-wave
electronics were intensively  discussed (see reviews~ \cite{Rev8,Knap}).
More recent  interest to the hydrodynamic phenomena in  low-dimensional
transport and plasmonics is driven by the
emergence of the high-mobility nanostructures \cite{Jaggi,
h1,h2,h3,h4,h5,h6,h7,h8} and graphene \cite{h9,h10,h11,h12,h13,h14,Levitov2016,Falkovich2017,
Danz2019,Xie2019,Titov2019} where
the electron-electron collision-dominated transport regime can be reached.

Two issues that have been most actively
discussed in recent years
are the emergence of hydrodynamic regimes with nonzero vorticity
(and their manifestation in the transport properties of the 2DEL) (see \cite{h7,Levitov2016,Falkovich2017,
Danz2019,Xie2019,Titov2019} and references therein),
as well as possible methods for measuring the electron viscosity  by using dynamic
excitations of 2DEL
\cite{h7}, and by nonlocal resistance measurements  \cite{Levitov2016,Falkovich2017,
Danz2019,Xie2019,Titov2019}.

Here, we demonstrate   that the electron flow with nonzero vorticity
can be excited by circular polarized  radiation. Importantly, we find that  such states appear even
in an ideal 2DEL with zero viscosity. We also find that the main effect  of viscosity  is broadening of
the plasmonic resonances in the structure shown in Fig.~\ref{Fig1}b.
Corresponding contribution  to the resonance width is proportional to  the kinematic viscosity
and  depends  on  the single geometrical factor---the distance $d$  between nanospheres.
This enables optical measurements of the electron liquid viscosity.

{\section{Model}}
{\subsection{Basic equations}}
In this work, we consider circular (twisted) plasmon excitation through the periodic array  of metal objects (or semiconducting objects with high conductivity),
such as
nanospheres or nanorings, embedded into or placed in the vicinity of the 2DEL
by using  insulating matrix transparent for the THz
radiation. To begin with, we consider the excitation by a single
nanosphere (see Fig.~ \ref{Fig1} a), and then  generalize the results in
the case of the grating  plasmonic coupler consisting   of  the a periodic array of  nanospheres (see Fig.~\ref{Fig1} b).

 Circularly polarized electromagnetic radiation  induces
a rotating dipole potential  in the nanosphere. As a result, an inhomogeneous  field
is formed, which, in turn, acts on the 2DEL.
We will find the dc  response of the system. We  assume that: (i) electron-electron collisions prevail over
scattering by phonons and impurities;  (ii) the radiation wavelength is
much larger than the radius of the nanosphere, so that the electric field
of radiation is uniform; (iii) the system is gated.
First assumption allows us to use the hydrodynamic approximation.

The 2D electron liquid  is described by the hydrodynamic equations for
the dimensionless electron concentration $n= (N-N_{0})/N_0 $ and velocity $ \m v $:
\begin{align}
\label{N}&\frac{\partial n}{\partial t}+{\rm div}\left[(1+n)\mathbf{v}\right]=0,
\\
\label{v} &\frac{\partial\mathbf{v}}{\partial
t}+\left(\mathbf{v}\nabla\right)\mathbf{v}+\gamma\mathbf{v} +s^2 \nabla n -\nu \Delta \m v
=\frac{e\mathbf E}{m}.
\end{align}
Here $N_0$ is equilibrium concentration, $s$  is the plasma wave velocity,
$\gamma$ is the rate of the momentum relaxation,
$\omega$ is the radiation frequency,  $m$ is the electron mass, and $\nu$ is the kinematic
viscosity.
The  field acting in the 2D plane,  $\mathbf E= \m E_0 (t)+\m E_1 (t,\m r)
$  is  given by the sum of the homogeneous field of  circularly polarized incoming radiation,
$\m
E_0(t)=E_0(\cos\omega t, -\sin \omega t) = (E_0/2)(\m e_x- i\m e_y)\exp[-i
\omega t] +c.c.  $   and  the dipole
field
\begin{equation}
\m E_{1}(\m r ,t)=-e\nabla \frac{\mathbf{r}\mathbf p(t)}{(r^2+a^2)^{3/2}},
\end{equation}
where $\m p(t)=p(\cos{\omega t},-\sin{\omega t})$ and $e p=E_0 R^3$ is the dipole moment of
a  metallic
 nanosphere with radius $R.$   (Alternatively, one can use dielectric
 nanospheres with dielectric constant $\epsilon_R.$  Then, the dipole moment
 becomes  $e p=E_0 R^3 (\epsilon_R +\epsilon)/(\epsilon_R + 2 \epsilon) ,$ where  $\epsilon$ is the
 dielectric constant of the  transparent  embedding matrix \cite{comment2}.)  Here, we assume   that internal plasmonic frequency of the
 nanospheres is very large as compared  to characteristic
 frequencies of the problem, so that  spheres are fully polarized
 {(corresponding estimates are given in Section~\ref{estimates}).}    For a   lattice of the nanospheres,  one should  replace
 $\m E_{1}(\m r ,t) \to  \sum_{i}
  \m E_{1}(\m r - \m r_{i} ,t), $  where  summation is taken over the lattice nodes.
{\subsection{Rectification of the optical signal}}
The incoming radiation leads to the  oscillations of
the concentration and velocity, which are rectified due to the nonlinearity  of the hydrodynamic equations.  The small signal solution of hydrodynamic
equations Eqs.~\eqref{N} and \eqref{v} can be found perturbatively  by
expansion over $E_0$ up to the second order
\be n  \approx  \delta
n(t,\m r)  +
\overline{n}(\m r),
\quad \m v \approx \delta \m v(t,\m r)  +
\overline{\m
v}(\m r),
\nonumber
 \ee
 where $\delta n(t,\m r) \propto E_0$
 and
 $\delta \m v(t,\m r) \propto E_0$
    are
 oscillations of the concentration and velocity representing
linear response, and $ \overline{n}(\m r) \propto E_0^2 $ and
$ \overline{\m v}(\m r) \propto E_0^2$ are time-independent corrections arising
due to the rectification. We will see that the optically-induced
flow of the 2DEL with nonzero vorticity appears even in an ideal   liquid with zero viscosity.
Therefore, we will first  put $\nu=0$ and discuss the viscosity related effects at the end of the paper.
One of our main findings is that a finite viscosity leads to a very simple contribution
to the width of the plasmonic resonances
and could be extracted from the measurements of the  resonance width.

 Due to the rectification, the impinging radiation induces both a dc current $\m j_{\rm dc}$ and  a static electric potential $\phi_{\rm dc }.$
To find the
rectified    corrections $ \overline{n} (\m r)$ and $ \overline{\m v} (\m r) $  (squared-in-$E_0$) we average
Eqs.~\eqref{N} and \eqref{v} over time thus arriving at the following set
of the stationary equations
\begin{align}
\label{Nr}&{\rm div}\hspace{0.3mm} \overline{\m{v}} = - {\rm div}\hspace{0.3mm}  \m J_{1},
\\
&
\label{vr}
\gamma \overline{\m v}
+ s^2 \nabla \overline{n}=  \gamma \m J_{2}
\end{align}
with the rectified sources.
{(We neglect  terms  of the order $E_0^2$ oscillating
at the frequency $ 2\omega.$  Such terms  leads to negligible, on the order of $ E_0^4,$
corrections to  the
 circular dc current.)}
\be\m J_{1}=  \langle \delta n \delta \m v \rangle_t,
\quad \m J_{2}= - \frac{1}{\gamma}{\langle \left(\delta
\m v\nabla\right) \delta \m v \rangle_t}.
\label{J}
\ee
To find total  radiation-induced  dc current, $\m j_{ \rm dc},$
 one should sum $\overline{\m v}$  and the
rectified source $\m J_{1}.$  The radiation-induced potential,
$\phi_{\rm dc}$  which creates static electric field $E_{\rm dc}= -\nabla \phi_{\rm dc} $
is found from the condition $ e \nabla \phi_{\rm dc}/m= s^2 \nabla \overline{n}. $
Thus, we have the following set of equations for $\m j_{ \rm dc}$ and  $\phi_{\rm dc}.$
\begin{align}
&\m j_{\rm dc} (\m r)= N_0 \left[\overline{\m v}(\m r) + \m J_{1}(\m r) \right], \label{jdc}
\\
&e \phi_{\rm dc } (\m r)= {ms^2}~ \overline{n}(\m r). \label{Phidc}
\end{align}

Hence, the key steps of the calculation  are as follows. One should first linearize
hydrodynamic equations \eqref{N} and \eqref{v} and find the linear  response.  The next step is to
substitute thus found   $\delta n$
and $\delta  \m v$ into the  expressions for the  non-linear  sources given by  Eq.~\eqref{J},
perform the  time averaging and find $\m J _{1,2}.$
  Then, one should calculate   $\overline{n}$
and $\overline{\m v}$ by solving Eqs.~\eqref{Nr}, \eqref{vr},  and,
finally,  find $\m j_{\rm dc}$ and $\phi_{\rm dc}$ from Eqs.~\eqref{jdc} and
\eqref{Phidc}.
\vspace{0.5cm}
\section{Linear response:  Drude and plasmonic contributions}
Since electric field entering  right-hand side of Eq.~\ref{v}, has both homogeneous and
inhomogeneous contributions,  one can present  the velocity oscillations as the sum of the  homogeneous Drude excitation   and inhomogeneous dipole-induced  plasmonic term, while
\be \delta \m v =\delta \m v^{\rm
D} +\delta \m v^{\rm P}  . \label{f,D} \ee
Corrections to the concentration appear only due to the inhomogeneous perturbation, so that
$\delta n= \delta n^{\rm P}$.
As we  demonstrate below, the  presence  of these two types of the velocity excitations
  leads  to interference effects,  and, as a consequence, to the   Fano-like asymmetry of the resonances.

  Linearizing Eqs.~\eqref{N} and \eqref{v} and  writing
  $\delta n= \delta n_{\omega}(\m r) e^{-i \omega t} +c.c.,
  ~\delta \m v= \delta \m v_{\omega} (\m r) e^{-i \omega t} +c.c. ,$  after simple calculations (see Appendix \ref{Linear})
  we get
\begin{align}
&\delta n_\omega(\m r)=\Delta  Z(\m r),
\\
&\delta \m v_\omega(\mathbf r)= \underbrace{i \omega\nabla Z(\m r)} \limits_{\rm \delta \m v_{\omega}^{\rm P}}
+\underbrace{\!\frac{eE_0(\m e_x-i\m e_y)}{2 m (\gamma- i \omega)}} \limits_{\delta \m v_{\omega}^{\rm D}},
\end{align}
where, for the case of a single nanosphere
  \be Z(\m r)= - i 2 \pi l^2 \int{\frac{d^2q}{(2\pi)^2}\frac{e^{i\mathbf{q}\mathbf{r}}e^{-i
\varphi_\m q} e^{-q a }}{q^2-k^2}}.
\label{Z}
\ee
Here $e^{-i \varphi_\m q}=(q_x-iq_y)/q,$
\be l^2=\frac {e^2 p}{2ms^2},\label{l}\ee
 and
\begin{equation}
k=\frac{\sqrt{\omega(\omega+i\gamma)}}{s}=k_{0}+i{Q}.
\label{k}
\end{equation}
The  real and imaginary parts of $k,$  respectively, $k_0$ and $Q,$  have a physical meaning
of  the wave vector and the spatial decrement of the  optically excited plasma wave.     In what follows,
we assume  $\gamma
\ll \omega.$ Hence, $k\approx{(\omega+i{\gamma}/{2})}/{s},$ and,
consequently, $k_0 \approx \omega/s$, $Q \approx \gamma/2 s .$   As seen, the spatial decrement
of the wave is small
\be
Q \ll k_0.
\ee

For the case of square dipole  lattice with the lattice constant $d,$  Eq.~\eqref{Z}
is slightly modified by   the replacement {(see Appendix \ref{app:Lattice})}
$$\int \frac{d^2q}{(2\pi)^2} \to  \frac{1}{d^2} \sum \limits_\m q,  $$
where  wave vector $\m q$ runs over the  inverse lattice vectors
\be
\m q_{nm} =\frac{2\pi}{d} \left( n \m e_x + m \m e_y \right)
\label{qmn}
\ee

Since velocity is given by the sum of two terms [see Eq.~\eqref{f,D}], one
can split  both of the rectified  sources $\m J_{1,2}$  into two contributions---the plasmonic contribution
and the  mixed (plasmonic+Drude) contribution:  $$ \m J_{i}= \m
J_i^{\rm P}+\m J_i^{\rm M}~ (i=1,2),$$
where
\be
\label{J12P-defenition}
\begin{aligned}
&\!\!\m J^{\rm P}_1\!\!=\!\langle \delta n^{\rm P} \delta \m v^{\rm P} \rangle_t,
~ \!\m J^{\rm P}_2\!\!=\!-\frac{\langle \left(\delta
\m v^{\rm P}\nabla\right)\! \delta \m v^{\rm P} \rangle_t}{\gamma},
\\
&\!\!\m J^{\rm M}_1\!\!=\!\langle \delta n^{\rm P} \delta \m v^{\rm D} \rangle_t,
~ \!\m J^{\rm M}_2\!\!=\!-\frac{\langle \left(\delta
\m v^{\rm D}\nabla\right)\! \delta \m v^{\rm P} \rangle_t}{\gamma},
\end{aligned}
\ee
Equations Eq.~\eqref{Z} and \eqref{J12P-defenition}
allow us to clarify basic physics issues in more detail.  First of all, as seen, the
integral in the r.h.s. of Eq.~\eqref{Z}
contains a  pole in the denominator, which reflects  the   plasmonic resonance occurring when
$\omega$ is equal to the  frequency of the plasma wave with the wave vector $q$.    However,
the pole is smeared out due to the integration over
$ \m q.$  The situation is different for  a dipole lattice when the integration should
be replaced   with summation. For small $\gamma,$    the contributions
of  the  different terms in the sum are well separated and  can give   sharp plasmonic resonances.
The  resonance  condition,
\be \omega=\omega_{nm}=  (2\pi s/d)\sqrt{n^2 +m^2},  \label{wnm}\ee  is
satisfied for  several pairs  $(n,m).$ For example, the fundamental plasmonic resonance with the frequency
\be
\omega_0 =\frac{2\pi s}{d}, \label{omega0}
\ee
 corresponds to the sum  over $4$  pairs $(1,0),(-1,0),(0,1)$  and $(0,-1)$  yielding
\be
Z_0(\m r) \propto  \frac{1}{\omega_0^2 - \omega^2 -i \omega \gamma},
\ee
with the frequency-independent  coefficient of proportionality. Then,   rectified  dc currents
  has the resonance dependence  $ \m J_i^{\rm P} \propto  |Z(\m r)|^{2}, $
$ \m J_i^{\rm M} \propto Z(\m r). $ As a result, in the vicinity of the resonance,  the expression for
the circular dc current can be  approximately presented as follows
\be
\m j_{dc} \approx \frac{\boldsymbol{ \pi}(\m r) }{\Omega^2+ \Gamma^2/4}
  +\left[\frac{ \boldsymbol{ \mu} (\m r) }{\Omega +i \Gamma/2} + c.c. \right].
\label{jdc0-res}
\ee
where
\be
\Omega= \frac{\omega-\omega_0}{\omega_0},\quad \Gamma=\frac{\gamma}{\omega_0},
\ee
are, respectively, the dimensionless  detuning  and damping of the fundamental resonance, while
the terms proportional
 to   vectors  $\boldsymbol{\pi} (\m r)  $ and   $\boldsymbol{ \mu}(\m r) $
represent  the plasmonic
and mixed contributions, respectively
[exact expressions for these coefficients will be given below, see Eqs.~\eqref{PP}, and \eqref{MM}].
 Due to the interference of these terms, the resonance in
$\m j_{\rm dc}$  and $\phi_{\rm dc},$
has an asymmtric Fano-like shape. Interestingly enough,  the degree of asymmetry depends on the coordinate $\m r.$

\begin{figure}[h!]
\centerline{\includegraphics[width=0.5\textwidth]{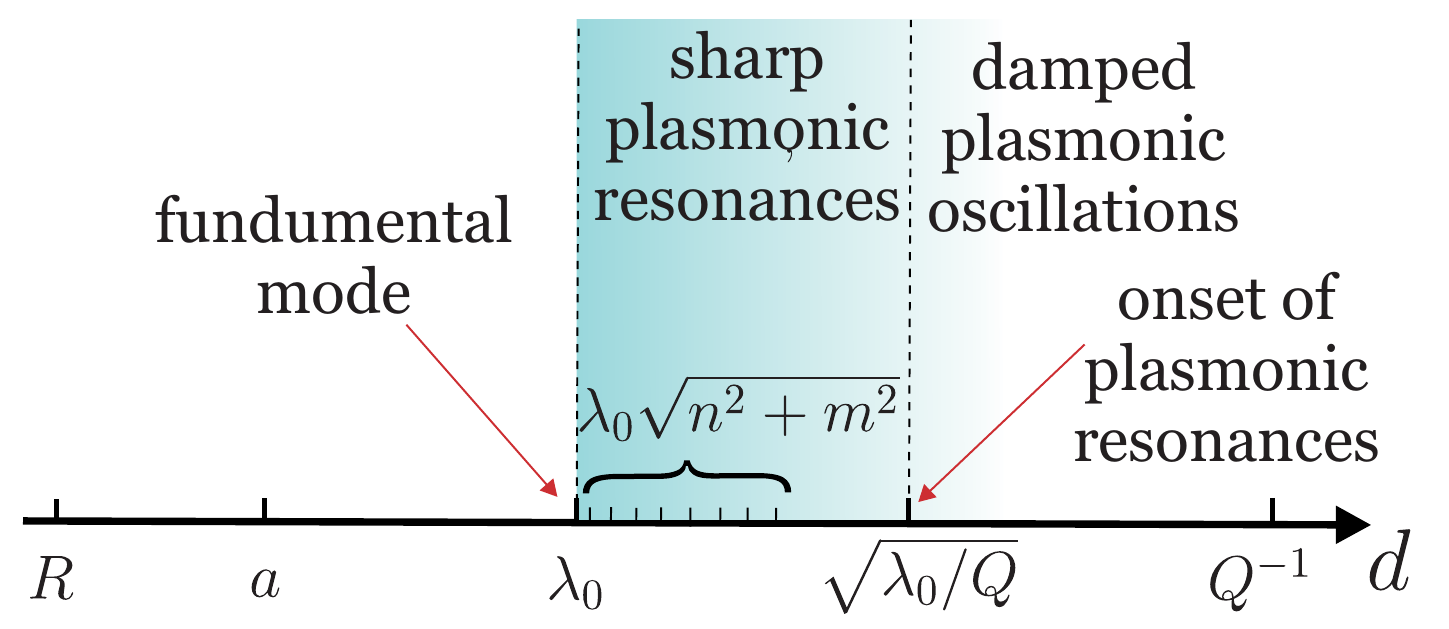}}
\caption{Different scales of the problem. We predict sharp plasmonic resonances  for
$ \lambda_0<d < \sqrt{\lambda_0/Q}.$    } \label{Fig_scales}
\end{figure}

 Different scales of the problem are illustrated in Fig.~\ref{Fig_scales}. The smallest scale is the size of
the sphere, $R,$ which is on the order or smaller than the distance from spheres to the plane of 2D gas,
$R\lesssim a.$   We assume that the wavelength of the plasma excitations, $\lambda_0=2\pi/k_0,$
is much larger than $a$ but smaller than the plasma wave   damping length:  $ a \ll \lambda_0 \ll Q^{-1}.$
For $d \gg Q^{-1},$ the spheres are fully independent  and it is sufficient to calculate the response
of a single sphere.  With decreasing $d,$ the spheres begin to influence each other. One can easily
 estimate characteristic
$d$  corresponding to onset of plasmonic resonances. To this end, we  estimate the
volume in the momentum space  corresponding to a plasmonic resonance  as $k_0 Q.$  When this volume
becomes smaller than the   volume of the unit cell   of the inverse lattice, $k_0 Q \ll  (2\pi/d)^2$
the resonances cease to   overlap.  The fundamental  mode corresponds to a smaller inter-sphere distance:
$d=\lambda_0.$  The total number of well resolved resonances
that can be observed is proportional to  $k_0 / Q = \omega/\gamma$  and is thus determined by the
quality factor.    It is worth noting that sharp  resonances exist in the finite range of $d$:
$ \lambda_0  \lesssim d   \lesssim\sqrt{\lambda_0/Q}.$

An important comment is related to the radiation-induced vorticity of the 2DEL.
On the formal level, function $Z(\m r)$ is a Green's function of hydrodynamic equations
describing     the plasmonic excitation caused  by  a  point-like rotating dipole.
Due to this rotation,  an  angular moment $ \pm 1$ is transferred to the liquid with the sign
 determined by the sign of the  helicity.  The information
about  this moment is encoded in  the phase factor $\exp[- i \varphi_\m q ]$ in Eq.~\eqref{Z}.
This means that  the plasma waves circulate
around  the nanospheres and that direction of circulation changes with changing the
sign of the radiation polarization.  We  call such excitations  ``twisted plasmons''. The rectification
 of these plasmons leads to dc current with non-zero vorticity, which is also determined by the helicity sign.

%
\begin{figure}[b]
\centerline{\includegraphics[width=0.5\textwidth]{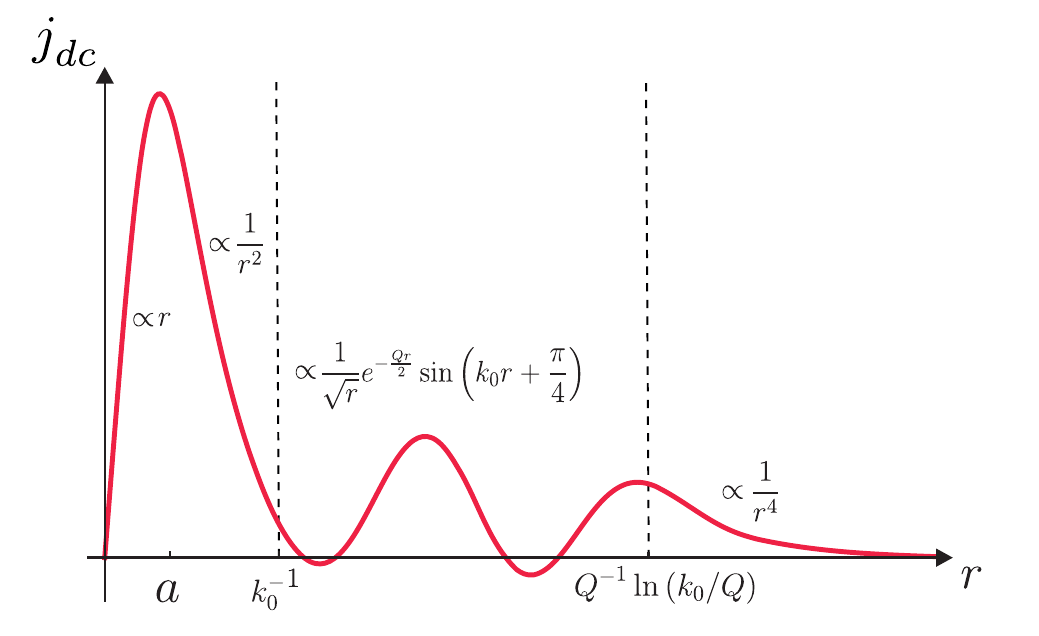}}
\caption{Dependence of the
circular current density, $j_{\rm dc}, $ created in 2D liquid   by a rotating dipole
moment  of a  single nanosphere. Main contribution to this current comes from mixed term [See Eq.~\eqref{jdc-mixed-single}]
 } \label{Fig2}
\end{figure}

\section{ Circular dc current induced by a single dipole}
Performing integration over $\varphi_\m q$ in Eq.~\eqref{Z}, we get
\begin{equation}
Z(\m r)=l^2(x-iy)f(r),
\label{Z-f}
\end{equation}
where function $f$  depends only on $r=|\m r|.$  The analytical expressions for $f$ and its
asymptotes
are   presented in the Appendix \ref{Linear} together
with expression of  $\delta n_\omega $ and $\delta v_\omega$ in terms of $f.$
It is convenient to present
$ \m J_{i}$ as follows \be \m J_{i}= R_{i} \m e_{\m r} +
\Phi_{i} \m e_{\m \phi},  \label{JP}\ee
where $\m e_{\m r} =\m r/r, $ $\m
e_{\m \varphi } = \m e_{z} \times \m e_{\m r} $
and  functions $R_{i}= R_{i}^{\rm P}+R_{i}^{\rm M},$ and $\Phi_{i}= \Phi_{i}^{\rm P}+\Phi_{i}^{\rm M},$
depend only on $r= |\m r|$ and contain both plasmonic and mixed contributions. {
Here,  vector $\m r $
is counted from the center of nanosphere (see Fig.~1a).}

 Provided that  $R_{i}$ and $\Phi_{i}$  are known,
  the solution of
Eqs.~\eqref{Nr} and \eqref{vr} can be found by expanding $\overline{\m v} $ over
$\m e_{\m r}  $  and  $\m e_{\m \varphi }$ and assuming $\overline{n}=\overline{n}(r).$ We
find for  the total circular radiation-induced
dc current, $\m j_{\rm dc} =  j_{\rm dc}  \m e
_{\varphi}$ and the  radial electric field, $  E_{\rm dc}= E_{\rm dc} \m e_{\m r}$:
\begin{align}
& j_{\rm dc}=N_0  (  \Phi_1 + \Phi_2)
\label{jdc12}
\\
&\frac{ e E_{\rm dc}}{m}= \gamma \left(R_1+ R_2 \right)
\label{Edc12}
\end{align}
Expressions for plasmonic and mixed contributions, $R_i^{\rm P}, \Phi_i^{\rm P}$
and $R_i^{\rm M}, \Phi_i^{\rm M},$ are presented in Appendixes \ref{app:JP} and \ref{app:JM},
respectively, as well as expressions for  asymptotical behavior  of
$j_{\rm dc}$ [see Eq.~\eqref{jdc-full}]  and $E_{\rm dc}$ [see Eq.~\eqref{Edc-full}]
accounting for both plasmonic and mixed contribution. As seen, for the most realistic
case ($R \ll a \ll k_0^{-1} \ll Q^{-1}$), the mixed contribution dominates. Neglecting
plasmonic contribution, we find that
\begin{widetext}
\be
j_{\rm dc} \approx - {j_*} \left\{
\begin{aligned}
& C\left (\frac{r}{a} \right), \quad  r \ll 1/k_0,
\\
&\frac{\sqrt{2\pi} k_0^{3/2} a^2 }{ \sqrt r} e^{-Qr/2} \sin (k_0 r + \pi/4), \quad 1/k_0  \ll  r \ll \frac{\ln\left[{k_0}/
{Q}  \right]}{Q}
\\
& \frac{6 a^2}{k_0^2 r^2},  \quad \frac{\ln\left[{k_0}/
{Q}  \right]}{Q} \ll r
\end{aligned}
\right.,
\label{jdc-mixed-single}
\ee
\end{widetext}
where
\be
j_*=\frac{\omega l^4 N_0}{ k_0^2 R^3 a^2},
\ee
and $C(x)$ is given by Eq.~\eqref{C}. Schematic dependence of $j_{\rm DC}$ on $r$ is shown in Fig.~\eqref{Fig2}. The static optically-induced field is  linked to the
dc circular current by a simple relation
\be
j_{\rm dc}= -\frac{e E_{\rm dc} N_0}{m \omega}.
\ee
\begin{figure}[h!]
\centerline{\includegraphics[width=0.4\textwidth]{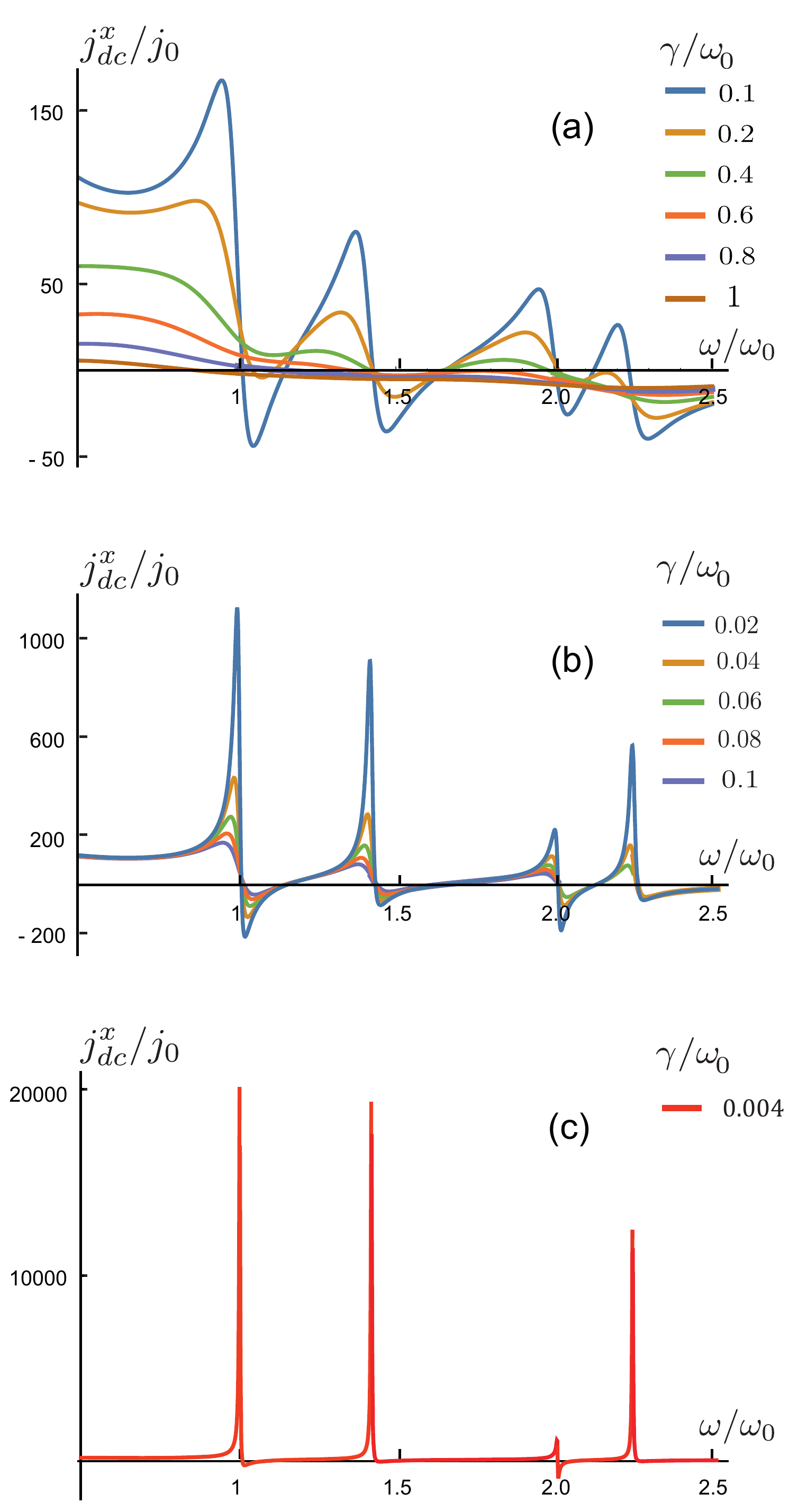}}
\caption{Frequency dependence of $x-$ component of the current density for $x=y=d/8,~R=a/2,~d=5a$:
 onset of plasmonic resonances  at large $\gamma$ (a); strongly asymmetric resonances
at intermediate values of $\gamma$ (b); weakly asymmetric resonances at very small $\gamma$ (c).
 } \label{Fig3}
\end{figure}
\subsection{Dipole lattice}
\begin{figure}[h]
\centerline{\includegraphics[width=0.4\textwidth]{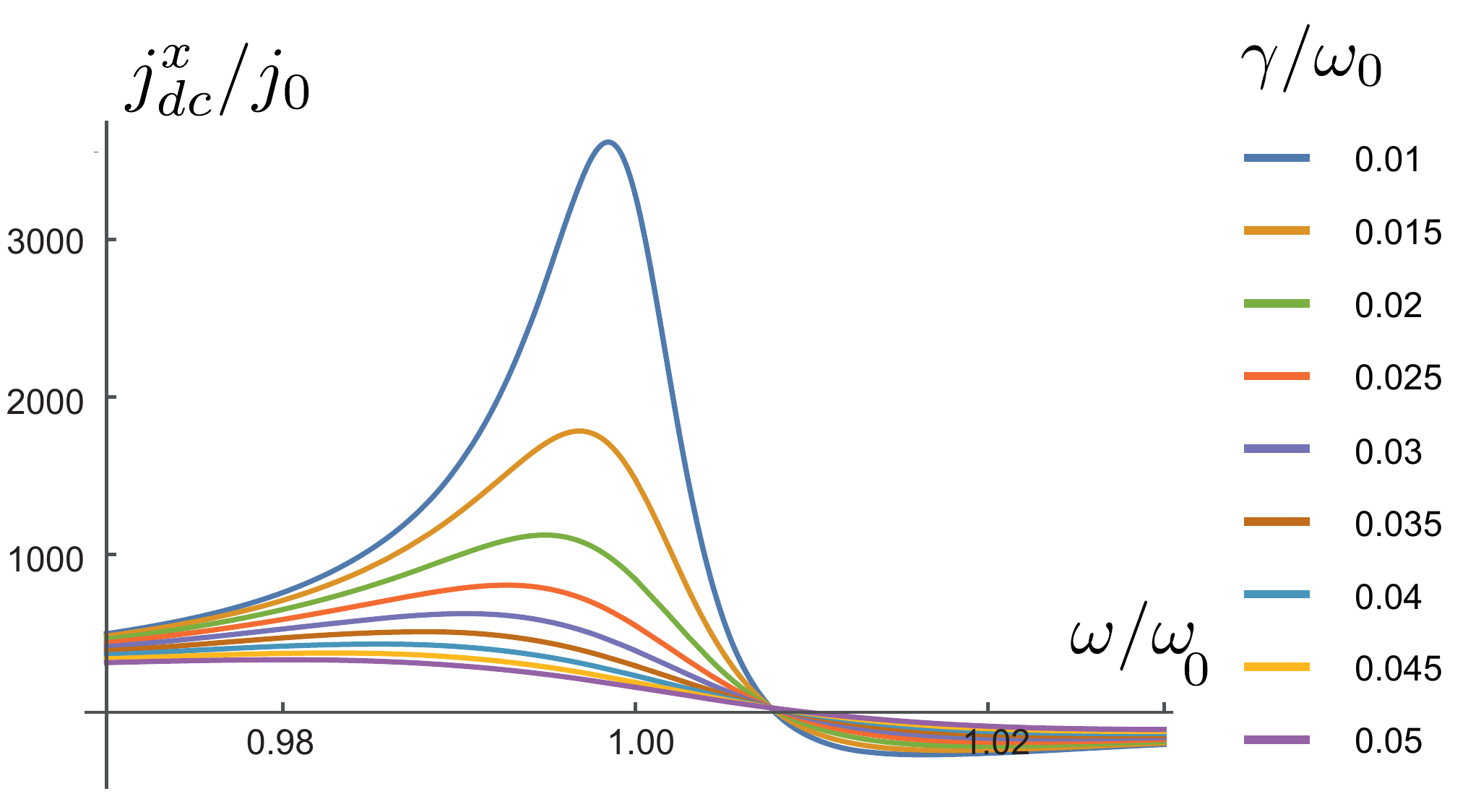}}
\caption{Fundamental plasmonic peak in $x$ component of dc current  for $x=y=d/8,$ $ R=a/2,$ $d= 5 a$
and different values of $\gamma.$   The asymmetry
of the peak decreases with decreasing of  $\gamma.$     } \label{Fig4}
\end{figure}
For a lattice of dipoles, we write the Fourier components of the nonlinear
sources $\m J_1$ and $\m J_2$ as follows
\be \m J_{i\m q\omega}= R_{i\m q}~  \m n_{\m q}^\parallel  + \Phi_{i\m q}~  \m n_{\m
q}^\perp,\quad i=(1,2)
\label{J-lattice-expansion}
\ee
where $\m n_{\m q}^{\parallel} = {\m q}/{q}$ and $\m n_{\m q}^\perp = \m e_z  \times {\m q}/{q}.$
 The Fourier transform of Eqs.~\eqref{Nr} and \eqref{vr} yields expressions similar to Eqs.~\eqref{jdc12} and
 Eq.~\eqref{Edc12}:
\begin{align}
& \m j^{{\rm dc}}_{ \m q }=N_0(\Phi_{1\m q} +\Phi_{2\m q})\m n_{\m q}^\perp,  \label{JDCQ}
\\
&\frac{ e \m E^{{\rm dc}}_ {\m q}}{m}=
\gamma \left( R_{1 \m q}+  R_{2 \m q} \right) \m n_{\m q}^{\parallel}. \label{EDCQ}
\end{align}
The Fourier components of the dc current and  static field can be presented
as sums over the plasmonic and mixed contributions: $ R_{i \m q}= R_{i \m q}^{\rm P}+ R_{i \m q}^{\rm M},$
$ \Phi_{i \m q}= \Phi_{i \m q}^{\rm P}+ \Phi_{i \m q}^{\rm M}.$

We consider the simplest case of a square  lattice with the lattice constant $d.$
In this case, all the integrals over $\m q$ should be replaced with the  sums over the
vectors of the inverse
 lattice [see Eq.~\eqref{qmn}] and
  function $Z(\m r)$ becomes
\be Z(\m r) = -\frac{i 2\pi l^2}{d^2} \sum \limits_{\m q} \frac{e^{ i \m q
\m r}}{ q^2-k^2} e^{- i \varphi_\m q } e^{- q a}. \ee Using this equation,
we find

\begin{align}
\delta n_\omega&= \frac{2 i\pi l^2}{d^2} \sum \limits_{\m q}
\frac{e^{ i \m q \m r}e^{- i \varphi_\m q } q^2 e^{- q a} }{ q^2-k^2},
\label{dnsum}
\\
\label{dvsum}
\delta \m v_\omega&= \frac{2 i\pi l^2}{d^2} \sum \limits_{\m q}
\frac{e^{ i \m q \m r}e^{- i \varphi_\m q }\omega \m q e^{- q a} }{q^2- k^2}
\\
&+\!\frac{eE_0(\m e_x-i\m e_y)}{2 m (\gamma- i \omega)}.
\nonumber
\end{align}

The rectified currents $\m J_{i}^{\rm P,M} $ can be calculated  using Eqs.~
\eqref{J12P-defenition},
\eqref{dnsum}, and \eqref{dvsum}.
Corresponding analytical expressions are given in Appendix \ref{app:Lattice}. Resulting equations for
$\m j_{ \rm dc}$ and $\m E_{\rm dc}$ are given, respectively, by Eqs.~\eqref{jdc-lattice} and \eqref{Edc-lattice}.

\begin{figure}[h]
\centerline{\includegraphics[width=0.3\textwidth]{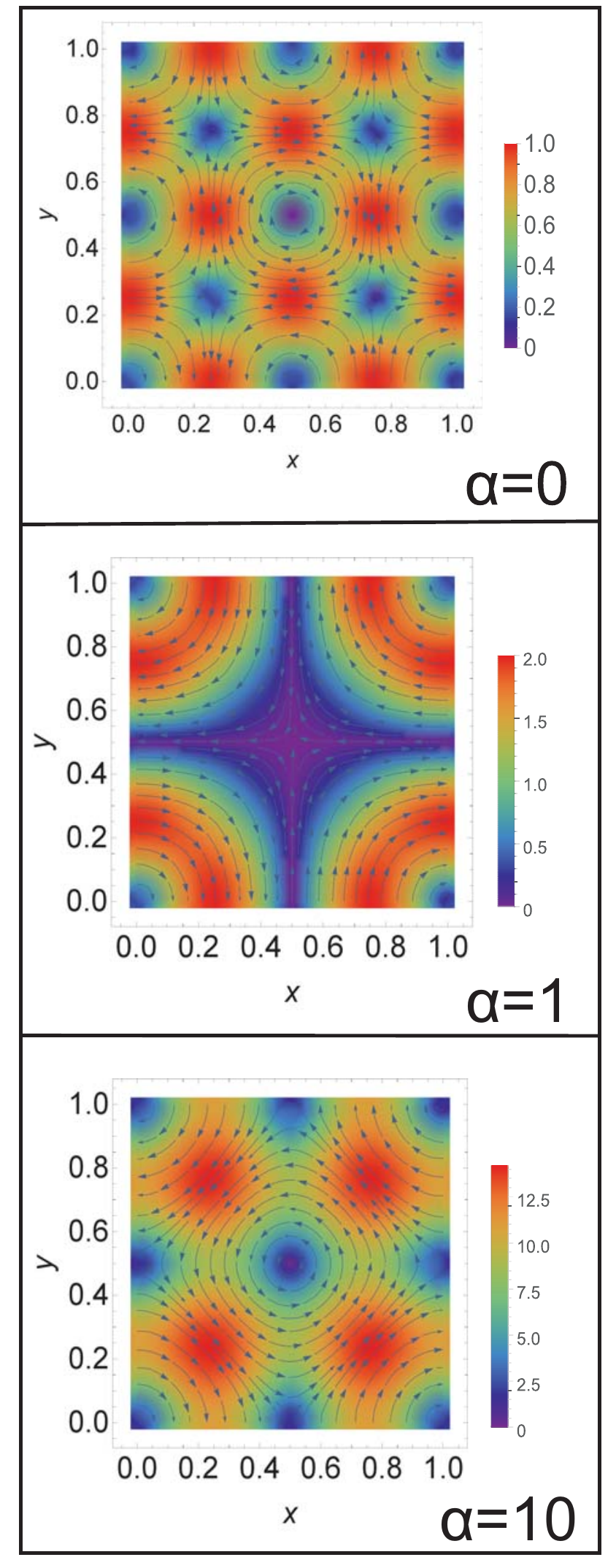}}
\caption{Vector density plot of the rectified current density
$\m j_{dc}$ for different values of parameter
$\alpha= 2\Omega \mu_0/\pi_0 .$  (here, $x$  and $y$ are measured in units of $d$)    } \label{Fig5}
\end{figure}

In  Fig.~\ref{Fig3}  we plotted the $x-$component of the dc current in units
of
$$j_{0} =N_0\frac{4\pi^2 l^4 s}{d^4},$$
in a certain point in
the plane (we used $x=y=d/8$) as a function of the radiation frequency for different damping
rates (picture for the $y$
component of the current looks analogous). As seen, with decreasing the $\gamma$,
sharp resonances appear on the top of the smooth dependence. Due to the
interference of the plasmonic and mixed contributions, the resonances have
an asymmetric  shape.
 The degree of asymmetry is smaller for small $\gamma$, because the
symmetric plasmonic contribution dominates at $\gamma \to 0$. Fig.~\ref{Fig4}
illustrates the asymmetry of the peaks for fundamental mode. To demonstrate vorticity of the
current, we also plotted the calculated current vector density in Fig.~\ref{Fig5}.

\subsection{Excitation of the fundamental mode}
{The smallest resonant frequency, $\omega_0,$   is given by Eq.~\eqref{omega0}.
This frequency corresponds to the contribution of four terms, with
\be
(n,m) = (1,0),(0,1),(-1,0),(0,-1).
\label{zeroM}
\ee
 For all these terms we have $q=q_0=2\pi/d.$
 The frequency of the  next resonance is given by $\sqrt 2 \omega_0.$
 It corresponds to  other four terms with   $(n,m) = ( \pm 1,\pm 1).$
 For sufficiently high quality factors, $$\omega_0/\gamma \gg 1,$$
these two resonances are well separated.  Hence, for  $\omega$ close to $\omega_0,$
only four terms corresponding to Eq.~\eqref{zeroM}
contribute to the
sum over $\m q_{nm},$ while
 terms with other $n$ and $m$ can be neglected (this corresponds to the resonance approximation).         }

{Within the resonance approximation,}  the concentration and velocity  are given by
\begin{align}
\delta n_\omega&= \frac{4\pi l^2}{d^2}~
\frac{ q_0^2 e^{- q_0 a}  [i \sin(q_0 y) -\sin(q_0x)]}{ q_0^2-k^2},
\label{ZM-dnsum}
\\
\nonumber
\delta \m v_\omega&\!=\! \frac{4 \pi l^2}{d^2}
\frac{\omega  q_0 e^{- q_0 a} [i \m e_x \cos(q_0 x)\! +\! \m e_y\cos(q_0y)]}{q_0^2- k^2}
\\ \label{ZM-dvsum}
&+\!\frac{eE_0(\m e_x-i\m e_y)}{2 m (\gamma- i \omega)},
\end{align}

Using the equations given in Appendix \ref{app:fund}, we find that  the circular current can be presented
in the form of Eq.~\eqref{jdc0-res}, with
\begin{align}
\label{PP}
\boldsymbol{ \pi}(\m r) & = \pi_0
\left [\sin (q_0y) \cos (q_0 x) \m e_x   \right.
\\
& \left. - \sin (q_0 x) \cos (q_0 y) \m e_y  \right], \nonumber
\\
 \boldsymbol{ \mu} (\m r) & = \mu_0 [\sin (q_0 y)  \m e_x - \sin (q_0x )  \m e_y],
\label{MM}
\end{align}
where
\be \label{pi0mu0}
   \pi_0=\frac{8\pi^2 N_0 s l^4}{d^4}e^{-2 q_0 a},~\mu_0= \frac{ N_0 s l^4}{d R^3}e^{- q_0 a}
  \ee
As seen, ${\rm div \boldsymbol{ \pi}}={\rm div \boldsymbol{ \mu}}=0,$ so that $\rm j_{\rm dc}$
is purely circular current, ${\rm div} \m j_{\rm dc} =0$, with non-zero vorticity:
\begin{align} \label{curl}
&{\nabla} \times  \m j_{\rm dc} =-\m e_z \frac{2 q_0 }{\Omega^2+ \Gamma^2/4}
\\
& \times \left  \{\pi_0 \cos (q_0x)\cos(q_0y)  \right. \nonumber
\\
&\left. + \mu_0 \Omega [\cos(q_0x)+\cos(q_0y)] \right \}.
\nonumber
\end{align}
Two interfering contributions, plasmonic and mixed, have different frequency dependencies
in the vicinity of  the resonance,
symmetric and asymmetric ones, respectively. Interestingly,
the degree of asymmetry depends on coordinate. For example, at the line $\cos(q_0x)+\cos(q_0y)=0$
the vorticity is a symmetric function of $\Omega,$ while for $\cos (q_0x)=0$  or $\cos(q_0y)=0,$
the vorticity is described by an asymmetric mixed term. The vector density plot of the rectified current
$j_{\rm  dc}$ is plotted in Fig.~\ref{Fig5} for different values of parameter
\be
\alpha= \frac{2\Omega \mu_0}{\pi_0}=\frac{d^3 \Omega}{4\pi^2 R^3} e^{q_0 a},
\ee
which depends on the dimensionless deviation from the resonance,  $\Omega. $ Hence, changing radiation
frequency, one can qualitatively change the spatial distribution of dc current.
{In order to understand this dependence better, we rewrite Eq.~\eqref{jdc0-res} as follows
\begin{align}
&\frac{\m j_{\rm dc}}{j_0} =\frac{2 e^{-2 q_0 a}}{\Omega^2+ (\Gamma/2)^2}
\left  \{ \m e_x \sin (q_0y)[\alpha+ \cos(q_0 x)] \right.
\nonumber
\\
&-  \left.\m e_y \sin (q_0 x)[\alpha+ \cos(q_0 y)] \right\}.
\label{currentZM}
\end{align}}
{
As seen, the key parameter which determines the current distribution is $\alpha.$
Below, we will analyze the vector structure of this equation. For
 brevity, we skip   common  coefficient  $2 j_0  { e^{-2 q_0 a}}/[{\Omega^2+ (\Gamma/2)^2}]$
 in the expressions for current.
}

{
 For $\alpha \ll 1,$ we get
 \begin{align}
 &\m j_{\rm dc}\!\propto\! \m e_x \sin (q_0 y) \cos(q_0 x)\! -\!
 \m e_y \sin (q_0 x) \cos(q_0 y)
\nonumber
 \\
 \nonumber
&j_{\rm dc}^2\propto \frac{1- \cos(2 q_0 x)\cos(2 q_0 y)}{2}.
 \end{align}
From these equations we find that the current reaches its maximum absolute
value at points   $ \m r^{\rm I}_{nm}= (x_n^{\rm I},y_m^{\rm I})= (d/2) \left( n+1/2,  m \right) $
and $\m r^{\rm II}_{nm}=(x_n^{\rm II},y_m^{\rm II})=
(d/2)\left( n,  m+1/2 \right)$ (here and below $n$ and $m$ are integer numbers). These points correspond
to centers of red circles in Fig.~\ref{Fig5}a.  From Eq.~\eqref{currentZM} we find values of
currents, $\m j_{\rm dc } ^{\rm I}$  and $\m j_{\rm dc } ^{\rm II},$
exactly at  $\m r^{\rm I}$ and  $\m r^{\rm II},$ respectively,
and their variations, $\delta \m j_{\rm dc } ^{\rm I}, \delta \m j_{\rm dc } ^{\rm II},$
in the vicinity of these points
\begin{align}
  &\m j_{\rm dc } ^{\rm I}  \propto -\m e_y (-1)^{n+m},
\nonumber
  \\
\nonumber
  &\delta \m j_{\rm dc }^{\rm I}  \propto (-1)^{n+m} q_0^2\left(\!
   \m e_y \frac{\delta x ^2 + \delta y^2}{2} -\m e_x \delta x \delta y\!\right),
\\
&\m j_{\rm dc } ^{\rm II}  \propto \m e_x (-1)^{n+m},
\nonumber
  \\
\nonumber
  &\delta \m j_{\rm dc }^{\rm II}  \propto (-1)^{n+m} q_0^2\left(\!
   -\m e_x \frac{\delta x ^2 + \delta y^2}{2} +\m e_y \delta x \delta y\!\right).
  \end{align}
Here, $\delta \m r =(\delta x,\delta y)$  is a small deviation of $\m r$ from point  $\m r^{\rm I}$
or $\m r^{\rm II}$ ($q_0\delta r \ll 1$). Analyzing these equations and Fig.~\ref{Fig5}a, we see that there are 8 current
maxima (per unit cell  of arising periodic structure) with different current behavior.
Here $\delta x $  and $\delta y$ are counted from  $\m r^{\rm I}$ or  $\m r^{\rm II}.$
For $\alpha \gg 1,$ we find that the current is given by
\begin{align}
&\m j_{\rm dc} \propto
\alpha \left[ \m e_x \sin (q_0y)+ \m e_y \sin (q_0 x)\right],
\nonumber
\\ &j_{\rm dc}^2 \propto
 \alpha^2 \left[\sin^2(q_0y)+  \sin^2 (q_0 x) \right].
\nonumber
\end{align}
We see  that dependencies on $x$ and $y$ fully decouple.
The current is maximal at points $(x_n,y_m)=  (d/2) (n+1/2,m+1/2), $  corresponding to
centers of the red circles  in Fig.~\ref{Fig5}c.
Close to these points, we get
\begin{align}
\nonumber
\m j_{\rm dc} &\propto (-1)^m  \alpha~ \m e_x \left( 1- \frac{q_0^2 \delta y^2}{2} \right)
\\
&+(-1)^n  \alpha ~\m e_y \left( 1- \frac{q_0^2 \delta x^2}{2} \right).
\label{max-new}
\end{align}
Hence, in this case there are four maxima with different current behavior
per unit cell of the periodic current structure.}

{The vector density plots for $\alpha \ll 1$ and $\alpha \gg 1$ are essentially different.
The transition between  these plots happens at $\alpha \sim 1.$
Let us consider, for example, the quadrant   of the unit cell of the  periodic structure of the current,
corresponding to $ 0<x<d/2$ and $0<y<d/2$ (behavior in the remaining three quadrants can be
considered analogously).   For $\alpha  =0,$  $j_{\rm dc}^2$ has four  maxima
of   equal heights at   the  points  $\m r^{\rm II}_{00},\m r^{\rm I}_{00},\m r^{\rm I}_{01} $
and $\m r^{\rm II}_{10} $
[see      Fig.~\ref{Fig5}a]. With increasing $\alpha$ first two maxima increase  by a factor
 $ (1+\alpha)^2, $ while the second two decrease by a factor $(1-\alpha)^2.$ Also,  for
 $\alpha < \sqrt{8/3}$ there is a
 saddle point in this quadrant  at
 \be
 x=y= \frac{d}{2\pi} \arccos \left(\frac{2}{\alpha+ \sqrt{\alpha^2+8}}\right).
 \ee
The squared current at the saddle  point  is
\be j_{\rm dc}^2 \propto\frac{4 (2 + a (a + \sqrt{8 + a^2}))^3}{(a + \sqrt{8 + a^2})^4}.
\label{saddle}\ee
At $\alpha > \sqrt{8/3},$ the saddle point transforms to a maximum and the amplitude of this
maximum  becomes higher than for the maxima at points $\m r^{\rm II}_{00},\m r^{\rm I}_{00}.$
With a further increase of $\alpha$  the new maximum moves  to the point $(x,y)=(d/4,d/4), $
and stops at this position for $\alpha \to \infty.$ The behavior of current in the vicinity
of this maximum at $\alpha \gg 1 $
 is described by Eq.~\eqref{max-new}. It is also  worth   noticing that for $\alpha \sim 1 $
 the value given by Eq.~\eqref{saddle} is close to the value of  maxima at
 points $\m r^{\rm II}_{00},\m r^{\rm I}_{00}.$
Therefore, the vector density plot shows red circular band (see Fig.~\ref{Fig5}b).
}

Analogously, one can  calculate the optically-induced  static potential
\begin{align}\label{phi-dc-final}
   & \frac{e \phi_{\rm dc}}{m} =\frac{2\pi^2 l^4 s^2}{d^4} \frac{e^{-2 q_0 a}}{\Omega^2+\Gamma^2/4}
    \\ \nonumber
   &  \times \left\{     \cos(2 q_0 x) +\cos(2 q_0 y)
   \right.
   \\ \nonumber
   & \left.
   - 4 \alpha \left[ \cos( q_0 x) +\cos( q_0 y)\right]  \right\}
.\end{align}
{As one can see from this equation, the maximal optically induced
voltage drop across different points of the unit cell  of the
periodic voltage structures is    proportional to the
amplitude of the circulating dc current       $e \delta \phi_{\rm dc}^{\rm max}
\sim   j_{\rm dc} s/N_0.$  }
\subsection{Optically-induced magnetic field}
The  stationary  radiation-induced magnetic field  obeys
\be
\left[ \nabla \times \m H\right]= \frac{4\pi e \m j_{\rm dc}(\m r)}{c}\delta(z).
\ee
Substituting  $\m H=\left[ \nabla \times  \m A\right]$ (${\rm div} \m A=0$) and making Fourier transform over
$\m r$, we find
\be
k^2 \m A_{\m k}- \frac{d^2 \m A_{\m k}}{d z^2}= \frac{4 \pi e \m j_{\rm dc}^{\m k}}{c} \delta(z).
\ee
Finite at $|z| \to \infty $ solution of this equation reads
$\m A_ \m k (z)= ({2\pi e}/{c k})\m j_{\rm dc}^{\m k} \exp(-k |z|).$
Hence, the Fourier transform of the  vector potential (and, consequently, of the  magnetic field) is  proportional to the
 Fourier  transform  of the dc  current. In the vicinity of  plasmonic peaks, only several $\m k$
 satisfying resonant  conditions contribute to the current  and magnetic field,
 so that spatial dependence of the field is found by the summation  over these discrete set of  $\m k.$

\begin{figure}[h]
\centerline{\includegraphics[width=0.3\textwidth]{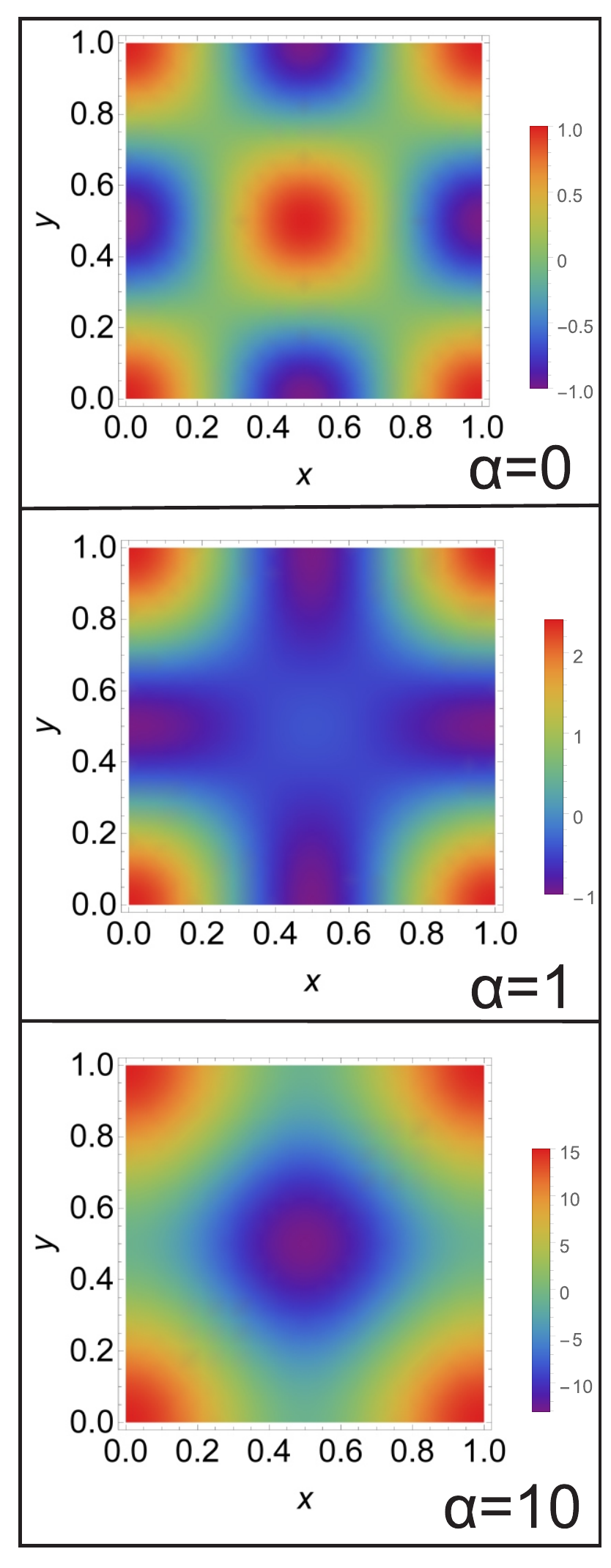}}
\caption{Density plot of $H_z(\m r,0)$ for different values of parameter
$\alpha= 2\Omega \mu_0/\pi_0 $ (here, $x$  and $y$ are measured in units of $d$)        } \label{Fig6}
\end{figure}

 Let us, for example,   calculate the  perpendicular component of the
 field, $H_z,$   in the fundamental mode  within the resonance approximation.
 In this case, $\m k$ runs over  $( \pm q_0,\pm q_0)$ for the plasmonic contribution and
 over $( \pm q_0,0)$ and $( 0,\pm q_0)$ for the mixed contribution [see Eqs.~\eqref{PP} and
 \eqref{MM}].
 Instead of summation over these $\m k,$ one can take into account that all terms  in $\pi (\m r) $
 and $\mu (\m r)$ are eigenfunctions of the Laplace operator, $\Delta,$ and
 present the  field in the operator form
 as
 \be
H_z(\m r,z) =
 \frac{e^{-\sqrt{-\Delta} |z| }} {\sqrt{-\Delta}  }
  \frac{2\pi e \m e_z [\nabla \times \m j_{\rm dc}(\m r )]}{c}.
 \ee
From this equation and Eq.~\eqref{curl}, we find
\begin{align} \label{curl1}
&H_z(\m r,z) =- \frac{4\pi e}{c}\frac{1 }{\Omega^2+ \Gamma^2/4}
\\
& \times \left  \{\pi_0 \cos (q_0x)\cos(q_0y) \frac{e^{-\sqrt 2 q_0 |z|}}{\sqrt 2}  \right. \nonumber
\\
&\left. + \mu_0 \Omega [\cos(q_0x)+\cos(q_0y)] e^{-\sqrt  q_0 |z|} \right \}.
\nonumber
\end{align}
Figure ~\eqref{Fig6} shows the density plot of the magnetic field in the 2DEL plane:
{
\begin{align}
&H_z(\m r,0)\! =\!- \frac{\sqrt{32}\pi e j_0}{c}\frac{ e^{-2 q_0 a} }{\Omega^2+ \Gamma^2/4}
\!\left  \{ \cos (\! q_0x\!)\cos(\!q_0y\!)   \right. \nonumber
\\
&\left. + \frac{\alpha}{\sqrt 2} \left[\cos(q_0x)+\cos(q_0y)\right]  \right \}.
\label{curlz=0}
\end{align}
For $\alpha \ll 1,$  the field has maxima (within the area   $0<x<d,$ $ 0<y<d$)     at the points
$ (0,0), (d,0),(0,d), (d/2,d/2), (d,d)  $
where $\cos(q_0 x) \cos(q_0 y) $ is maximal (these maxima have equal heights and correspond
to centers of red circles   in   Fig.~\ref{Fig6}a).
With increasing $\alpha,$ the amplitude of the  central  maximum  at $(d/2,d/2)$ decreases by the factor
$1-2\alpha $ (for $\alpha>1$ this maximum transforms into minimum), while the
amplitude of other four maxima is increased by the factor $1+2 \alpha.$
 Hence, for  $\alpha \gg 1, $ the field has four equivalent maxima  at the points
 $ (0,0), (d,0),(0,d), (d,d) $ corresponding to the
maxima  of  both $\cos(q_0x) +\cos(q_0 y) $  and  $\cos(q_0 x) \cos(q_0 y) $  (the maxima   correspond
to centers of red circles in Fig.~\ref{Fig6}c).  }

{\section{Discussion }}
{\subsection{Finite viscosity, external magnetic field  and finite size effects.}} Above, we presented calculations for zero external  magnetic field for an ideal infinite   2DEL  with zero viscosity.
The detailed analysis of  different  magnetoresponse regimes of  the  viscous  electron liquid
in the system  under discussion  is out of scope of this work and will be presented elsewhere. Here, we limit
ourselves to the simplest but at the same time the most interesting case of the resonant  excitation,
when some of the  plasmonic modes with wavevectors given by Eq.~\eqref{qmn}
satisfy the resonance condition: $\omega \approx \omega_{nm},$ where $\omega_{nm}$
is given by Eq.~\eqref{wnm}. In this case, within the resonant approximation, the effect of
a weak magnetic field, $B,$  with   $\omega_c  \ll \omega$ ($\omega_c=eB/mc$ is the cyclotron frequency)
can be   accounted for by replacing $\omega_{nm}^2$ with
\be
\omega_{nm}^2(B) = \omega_{nm}^2 +\omega_c^2.
\ee
Hence, a  weak magnetic field shifts the positions of the resonances shown in Fig.~\ref{Fig3}
thus giving an  additional way to control the  dc current and magnetization.

Within the same resonance approximation,  the effect of a weak viscosity,  satifying the inequalty
$\nu q_{nm}^2 \ll \omega,$  is accounted by   replacing elastic damping $\gamma$ with
\be
\gamma_{nm} =\gamma + \nu q_{nm}^2.
\label{gamma-nm}
\ee
The resonance is described by Eq.~\eqref{jdc0-res} with
\be
\Omega \approx \frac{\omega-\omega_{nm}(B)}{\omega_{nm}(B)},
\quad \Gamma \approx \frac{\gamma_{nm}}{\omega_{nm}}.
\ee
As seen from Eq.~\eqref{gamma-nm}, the
measurement of widths  of two plasmonic resonances with different resonance
frequencies ($\omega_{n_1m1} \neq \omega_{n_2m_2}$)  allows  one to extract value of $\nu$:
\be
\nu=\frac{(\gamma_{n_1m_1} -\gamma_{n_2 m_2}) d^2}{(2\pi)^2 (n_1^2+m_1^2- n_2^2-m_2 ^2)}.
\label{nu}
\ee
Evidently, one can also extract the momentum relaxation time by measuring $\gamma_{n_1m_1}$
  and $\gamma_{n_2 m_2}.$
It worth noting that Eq.~\eqref{nu}  does not include any  characteristic of the material
and depends on a single  geometrical factor---the distance between nanospheres,
which can be well controlled in experiment.
Hence,  the HIFE gives a direct way to extract the electron viscosity.

 In this paper, we considered an infinite 2D system. An
interesting question is related to finite size  effects, i.e. to the
behavior of the current and magnetic field at the boundary of the system. A
detailed discussion of this issue is beyond the scope of this work and
will be studied elsewhere. Here we restrict ourselves to a few
comments. One can consider the situation when a diffraction square lattice
having a finite size $ L = N d, $ (here $ N \gg 1 $ is an integer number)
is located over an infinite 2D plane. Then, when calculating the function
$ Z(\m r) $ in integrals over $ d q_i $ ($ i = x, y $), the factors $~  \sin (q_i L/2) /
\sin (q_i d/2) $ appear, which describe the smearing of $ \m q $ around the
quantized vectors of inverse lattice of $ \m q_{nm} $ [see Eq.~\eqref{qmn}] by values  of the order of $ \delta
q_i \sim 1 / L. $   Considering the fundamental  mode and calculating the corresponding
integrals, one can show that  outside of the region covered by diffraction lattice the plasmonic and mixed
contributions exponentially decay with different exponents
\be
\pi(\m r)\! \propto \exp\left(-{\delta r }/{L_{\pi}}\right),
\quad \mu(\m r)\! \propto \exp\left(-{\delta r }/{L_{\mu}}\right),
\nonumber
\ee
 where $\delta r $ is distance  from  the edge  of  the diffraction region,
$L_\pi = 1/\sqrt{k_0 Q} = s/\sqrt{\omega_0 \gamma} ,$ and  $L_\mu = 1/Q= s/ \gamma .$
It is worth noting
that for a small damping rate both  $L_\pi$ and  $L_\mu$ might become  on the order or even
larger than $L,$ which means that for sufficiently clean 2DEL the  circular current and magnetic filed
can appear well beyond the region covered  by diffraction lattice.
\vspace{0.5cm}
{\subsection{Estimates of relevant parameters for various structures}\label{estimates}}
\vspace{-0.5cm}
 {In this Section, we present some estimates of the relevant physical parameters
 for various materials and briefly discuss applicability of our approximations for realistic structures.
We  use  the following geometrical parameters: $d=250$ nm, $a=50$ nm, $R=25$ nm. The plasma wave velocity is estimated by using standard equation \cite{Dyakonov93} and assuming that
there is the back gate in the system. The  barrier (spacer) width given in Table~\ref{tab1}
corresponds to the typical values for each material system.
The electric field of the incoming radiation is taken as $E_0=10^5$ V/cm.
We estimate both current $j_0,$ which characterize the current flow for non-resonant case,
 when the   quality factor is on the order of unity,
and also  the current
\be
j^{\rm m}_{\rm dc } = \frac{ 8 j_0}{\Gamma^2} \exp^{-4\pi a/d},
\ee
which is much larger than  $j_0$ for sharp resonances, when $\Gamma \ll 1$ [see Eq.~\eqref{currentZM}].
For estimates of  plasmonic-enhanced magnetic field we use
\be
H^{\rm m}=\frac{16\sqrt{2}\pi e j_0 }{c \Gamma^2}\exp^{-4\pi a/d},
\ee
[see Eq.~\eqref{curlz=0}].      }

 {Table~\ref{tab1}   list the calculated values of the most important parameters,
 i.e. fundamental frequency, quality factor,
 characteristic value of the dc current, and maximal  magnetic field. For all the materials listed
  in this Table, the frequency of the fundamental plasmonic mode is in
  the THz range.   The value of optically-induced magnetic field can be sufficiently
 large at  not too low temperature, 77 K, especially in the GaN and p-diamond-based structures.     }
%
{For these estimates  we used  material parameters  listed in Table~\ref{tab2}
with references to corresponding experiments and/or numerical simulations.
Using the numbers, presented in the above Tables, we can discuss validity of approximations
used in our  calculations.}

\begin{figure}[h]
\centerline{\includegraphics[width=0.5\textwidth]{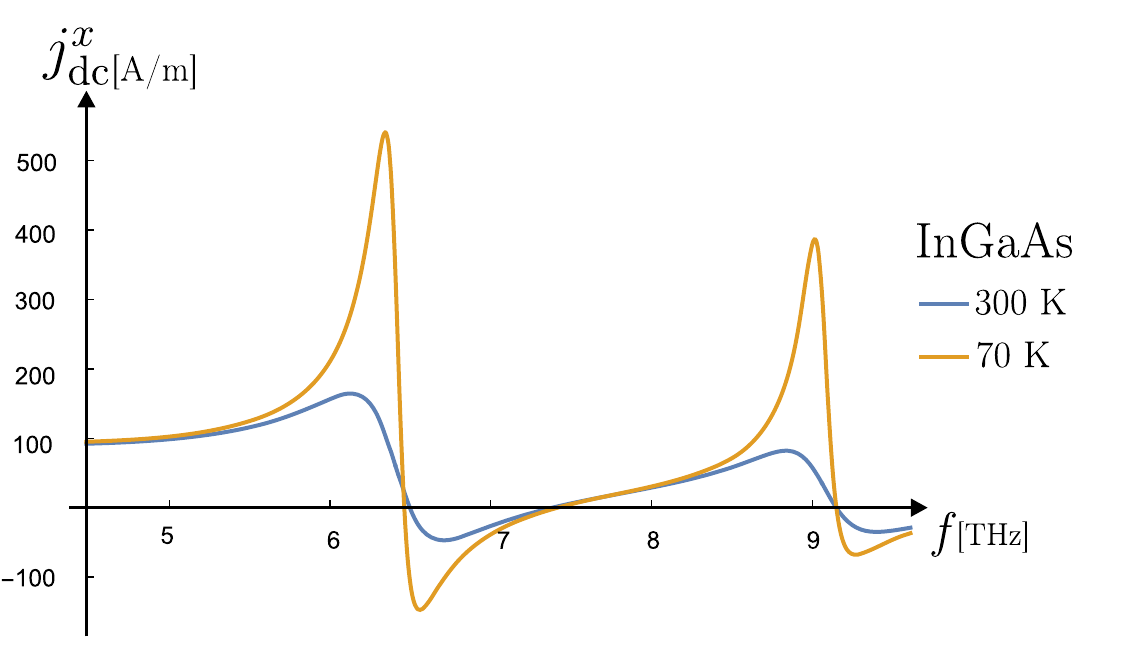}}
\caption{  Current-frequency dependence  for parameters of InGaAs-based structure
 with $d= 5 a=10 R= 250$ nm at $x=y=d/8$,    for two temperatures, 300 and 77 K.    } \label{Fig7}
\end{figure}

{In our model  we assumed that the  spheres  comprising the plasmonic coupler
 are fully polarized. This implies     that internal plasmonic frequency of the
 nanospheres is very large  compared  to characteristic
 frequencies of the problem.  The condition is well
 satisfied provided that  frequency of three dimensional  plasma
 oscillations in the metal, which the spheres
 are made from,
 $\omega_{\rm 3D},$  is large as compared to typical plasmonic frequency in our problem, which is the
 fundamental frequency   $\omega_0$ [see Eq.~\eqref{omega0}]. For typical value of plasma
 wave velocity in 2D gated  InGaAS-based  structure,
 $s \sim 1.6 \times 10^8$ cm/s  \cite{Dyakonov93}, and $d=250$ nm, we get
 $f_0 = \omega_0/2\pi=6.4$ THz (see Table~\ref{tab1} and Fig.~\ref{Fig7}).
 At the same time the 3D  plasmonic frequencies in metal are at least two order of
 magnitude  higher due to very high electron
 concentration.  For example, simple estimate  for silver, with
 3D concentration  $6 \times 10^{22}$ cm$^{-3},$
 yields  for $\omega_{\rm 3D} $ value about $10^{16}$ s$^{-1}.$}

{Let us now estimate spatial scales shown in Fig.~\ref{Fig_scales}.  Assuming the  frequency
of the radiation to
be $\omega = 3\times 10^{13}~ {\text s}^{-1}$ (which corresponds to $f=\omega/2\pi=5$ THz),
 and using estimate for typical plasma wave velocity $1.6\times10^8$ cm/s  we find  $\lambda_0=250$ nm.
 Rewriting damping length as
 $Q^{-1}= \lambda_0 (\omega_0/\gamma \pi)$  and using data of Table~\ref{tab1} for InGaAs at $T$=77 K,
 we estimate  $Q^{-1} \approx 3500$ nm.  This justifies ordering of the   spatial scales  in
 Fig.~\ref{Fig_scales}.
As seen from Table~\ref{tab1}, for  other materials we also have  $Q^{-1} \gg \lambda_0.$}

 {Finally, we present a picture of the current for parameters of InGaAs-based structure
 with $d= 5 a=10 R= 250$ nm,    for two temperatures, 300 and 77 K
 (see Fig.~\ref{Fig7}). We see  resonance at the  fundamental frequency
 $f=f_0=\omega_0/2\pi$ and the next one at the frequency  $\sqrt {2} f_0.$
 As expected,  the quality factor of plasmonic resonances increases with decreasing
 the temperature.      }

\begin{widetext}
\begin{table}[h]
	\caption{
		The estimated parameters for different structures
	}
	\begin{tabular}{|p{1.6cm}|p{1.3cm}|p{0.8cm}|p{1.2cm}|p{1.3cm}|p{1.3cm}|p{1cm}|c|p{1 cm}|p{1.5cm}|}
		\hline
		&Barrier thickness (nm)& T(k)& 2D carrier density $(1/cm^2)$&$f$=$\omega_{0}$/2$\pi$  (THz)&$\gamma/\omega_{0}$&$ej_0$ (A/m) &$ j_{\rm dc}^{\rm m}$/$j_{0}$&$ej_{\rm dc}
^{\rm{m}}$ (A/m)& $H^{\rm m}$(Gs)\\
		\hline
		GaN&20&300&$10^{13}$&7&0.087&0.1&87&9&0.086\\
		\hline
		GaN&20&77&$10^{13}$&7&0.005&0.1&21897&2299&20.4\\
		\hline
		Si&4&300&$2\cdot 10^{12}$&2.3&0.4&0.84&3.5&3&0.025\\
		\hline
		Si&4&77&$2\cdot 10^{12}$&2.3&0.03&0.84&657&550&4.9\\
		\hline
		InGaAs&20&300&$2 \cdot 10^{12}$&6.4&0.089&0.89&84&74&0.64\\
		\hline
		InGaAs&20&77&$2\cdot 10^{12}$&6.4&0.03&0.88&701&618&5.5\\
		\hline
		p-diamond&4&300&$2\cdot 10^{12}$&1&0.08&0.8&112&89&0.78\\
		\hline
		p-diamond&4&77&$2 \cdot 10^{12}$&1&0.01&0.8&4815&3814&34\\
		\hline
	\end{tabular}
	\label{tab1}
\end{table}
\begin{table}[h]
  \centering
\caption{Material parameters used in the calculations}
\begin{tabular}{|c|p{2.3cm}|p{2.3cm}|p{2.3cm}|p{2.4cm}|p{2.2cm}|}
	\hline
	& Effective mass&Mobility cm$^2$/Vs (77k)&Mobility cm$^2$/Vs (300k)&Dielectric constant of material&
Dielectric constant of barrier
\\
	\hline
	Silicon&0.19\cite{E8}&20000\cite{E6}&1450\cite{E5}&11.9\cite{E8}&3.9\\
	\hline
	GaN&0.23\cite{E10}&31700\cite{E1}&2000\cite{E2}&8.9\cite{E8}&8.9\\
	\hline
	InGaAs&0.041\cite{E9}&35000\cite{E7}&12000\cite{E7}&13.9\cite{E8}&12.1\\
	\hline
	p-diamond&0.663\cite{E4}&35000\cite{E3}&5300\cite{E3}&5.7\cite{E8}&5.7\\
	\hline
\end{tabular}
  \label{tab2}
\end{table}

\end{widetext}

\section{Conclusion}
To conclude, we predicted  excitation of circular plasmonic modes
(twisted plasmons)  in two dimensional
electron liquid  by circularly-polarized electromagnetic
wave via a plasmonic coupler made of periodically placed nanospheres.
We demonstrated that rectification of the
plasmons leads to a helicity-sensitive circular dc current, and
consequently, to a magnetic moment, thus demonstrating the hydrodynamic
inverse Faraday effect. This effect is dramatically increased in vicinity
of plasmonic resonances, so that the dc current shows sharp plasmonic
peaks. There are two interfering contributions to the peaks, the plasmonic
contribution, and the contribution involving both the  plasmonic and the
Drude excitations. As a result, plasmonic resonances  have  asymmetric  Fano-like shape.
 The
suggested system can be used for optical tunable magnetization of 2D
systems, for many optoelectronic devices operating in the THz range
of frequencies, and for the characterization and parameter extraction of
2D electron liquids. In particular, measuring  of the widths    of
different  plasmonic resonances allows one to extract the electron viscosity.
\section{Acknowledgement}
 The work of V.Yu.K.  was supported by RFBR (Grant No. 20-02-00490), by Foundation for the Advancement of Theoretical
Physics and Mathematics “BASIS,” and by the Foundation
for Polish Science through the Grant No. MAB/2018/9 for
CENTERA. The work at RPI was supported by the U.S. Army Research
Laboratory Cooperative Research Agreement (Project Monitor Dr. Meredith Reed)
and by the US ONR (Project Monitor Dr. Paul Maki).

\appendix
\begin{widetext}
\section{Linear  response (technical details) } \label{Linear}
Linearizing Eqs.~\eqref{N} and \eqref{v} and making Fourier
transform we get
\begin{align}
&-i\omega \delta n_{\omega \m q}+i\mathbf{q} \delta \mathbf{v}_{\omega \m q}=0, \label{eq_n_qw}
\\
&i\mathbf{q}s^{2} \delta n_{\omega \m q}\! +\!(\gamma-i\omega) \delta \mathbf{v}_{\omega \m q}
=\!
 \frac{e (\m E_0 + \m E_1)_{\omega \m q}}{m},\! \label{eq_v_qw}
\end{align}
where
\begin{align}
 &\left (\frac{e \m E_0}{m} \right)_{\omega \m q}= \frac{eE_0}{2m} (\m e_x - i \m e_y) (2
 \pi)^2\delta(\m q)
\\
&\left (\frac{e \m E_1}{m} \right)_{\omega \m q}= -\frac{\pi \m q e^2 p}{m} e^{- i \varphi_{\m q}}
 e^{- q a}
 \end{align}
and  $e^{-i\varphi_{\m q}}= (q_x- i q_y)/q$

Solution of Eqs.~\eqref{eq_n_qw}, \eqref{eq_v_qw}  reads
\begin{align}
&\delta n_{\omega\m q}= 2\pi i l^2 \frac{q^2}{q^2-k^2}    e^{-i \varphi_{\m q}} e^{-q a},
\\
&\delta \m  v_{\omega \m q}=2\pi i l^2 \frac{ \omega \mathbf {q}}{q^2-k^2}e^{-i \varphi_{\m q}} e^{-q a}
\\
&+\!\frac{eE_0(\m e_x-i\m e_y)}{2 m (\gamma- i \omega)} (2\pi)^2 \delta(\m q),
\nonumber
\end{align}
where $l$ and $k$ are given by Eqs.~\eqref{l} and \eqref{k} of the main text.
Next, we find the Fourier transform of the velocity and concentration:
\begin{align}
&\delta n_\omega(\m r)=\Delta  Z(\m r),   \label{dn-Z}
\\
&\delta \m v_\omega(\mathbf r)= i \omega\nabla Z(\m r) \label{dv-Z}
+\!\frac{eE_0(\m e_x-i\m e_y)}{2 m (\gamma- i \omega)},
\end{align}
where
\begin{align}
&Z(\m r)= - i 2 \pi l^2 \int{\frac{d^2q}{(2\pi)^2}\frac{e^{i\mathbf{q}\mathbf{r}}e^{-i
\varphi_\m q} e^{-q a }}{q^2-k^2}} \nonumber
\\
&=l^2(x-iy)f(r).
\end{align}
Function $f(r)$  is given by
\begin{align}
\label{f-struve}
&f(r)= \int_{0}^{\infty}\frac{dq q J_{1}(qr)e^{-qa}}{ r(q^2 -k^2)} \approx
\\&\!\frac{\pi}{2r}\!\!\left[\mathbb{ H}_{-1}(kr)\!+\!iJ_{1}(kr)\right]\!
-\!\frac{1}{r}\left(\!\! 1\!-\!\frac{r}{a\!+\!\sqrt{a^2+r^2}}\!\!\right),
\nonumber
\end{align}
where $\mathbb{H}_{-1}$ and $J_1$  are the Struve and Bessel functions.
Here we assumed  $Q \ll k_0  \ll 1/a$ \cite{comment1}.

The asymptotes of the  function $f$ are given by
\be
f \approx \left\{
\begin{aligned}
 &\sqrt{\frac{\pi}{2 k r^3}}e^{i(k r-\pi/4)} \left(1+\frac{3i}{8 k r} \right)
  -\frac{1}{k^2 r^3},  \qquad r  \gg  1/k_0,
  \\
 &\frac{1}{a+\sqrt{a^2+r^2}  } + \frac{i\pi k}{4}, \qquad r \ll 1/k_0
 \end{aligned}
\right.
\label{f-asym}
\ee
From Eqs.~\eqref{Z-f}, \eqref{dn-Z}, and \eqref{dv-Z} we get
\begin{align}\delta n_\omega^{\rm P}(\m r)
&=l^2 (x-i y)\left[f''+ \frac{3f'}{r}
\right] \label{dnf},
\\
\delta \m v_\omega^{\rm P}(\m r)\!&=\!\omega l^2 (x-i y)\!\!\left[ i \frac{(r f)'}{r}\m e_{r} \!+ \!
\frac{f}{r} \m e_{\varphi}\right],
\label{dvf}
\\
\delta \m v_\omega^{\rm D}(\m r)&=\!\frac{eE_0(\m e_x-i\m e_y)}{2 m (\gamma- i \omega)}.
\label{dvD}
\end{align}
As seen, the  velocity oscillations can be presented as a sum of the
$f-$dependent inhomogeneous contribution and  the  homogeneous Drude
contribution,
 given, respectively, by   Eq.~\eqref{dvf} and  Eq.~\eqref{dvD}

\section{Expressions  for  $R_i^{\rm P}$ and   $ \Phi_2^{\rm P}$ for   a single nanosphere}
\label{app:JP}
Using Eqs.~ \eqref{f-struve}, \eqref{f-asym},\eqref{dnf}, \eqref{dvf}, \eqref{dvD},  and \eqref{J12P-defenition},
 we find
\begin{align}
& R^{\rm P}_1 = -i \omega l^4 (r f''+ 3 f') (r f^*)' +c.c.
\approx \pi \omega l^4 \left\{\begin{aligned}
                                 &\frac{k_0^2}{r} e^{-Qr}, \quad r> 1/k_0, \\
                                 &\frac{k_0 r}{2 (r^2+a^3)^{3/2}}, \quad r<1/k_0,
                               \end{aligned}
  \right.
\label{R1P}
\\
&\Phi^{\rm P}_1 =  \omega l^4 (r f''+ 3 f') f^* +c.c.
\approx \pi \omega l^4 \left\{\begin{aligned}
                                 &-\frac{k_0}{r^2} e^{-Qr}, \quad r> 1/k_0, \\
                                 &-\frac{2}{\pi }\frac{ r}{ (r^2+a^2)^{3/2} (a + \sqrt{r^2+a^2})}, \quad r<1/k_0,
                               \end{aligned}
  \right.
  \label{Phi1P}
\\
&R^{\rm P}_2= - \frac{\omega^2 l^4}{\gamma}\left [ (r f^*)'(r f)'' +f^* f' \right] +c.c.
\nonumber
\\
&\approx  \frac{\pi \omega^2 l^4}{\gamma}
\!\left\{\begin{aligned}
                                &\frac{k_0}{2 r^2} (1+Q r)  e^{-Qr}, \quad r> 1/k_0, \\
                                &\!\! \frac{2}{\pi }\frac{ r}{  (a + \sqrt{r^2+a^2})^3}
                                 \!\!\left[\! \frac{a^3}{(a^2+r^2)^{2}}\!
                                 +\! \frac{r^2+3 a^2}{(a^2+r^2)^{3/2}}\!\right], \qquad r<1/k_0,
                               \end{aligned}
  \right.
  \label{R2P}
\\
& \Phi^{\rm P}_2 =0.
\label{Phi2P}
\end{align}
\section{Expressions  for   $R_i^{\rm M}$ and   $ \Phi_i^{\rm M}$ for   a single nanosphere}\label{app:JM}
By
direct averaging of $\delta n^{\rm P} \delta \m v^{\rm D}$ over time we get
\be R_1^{\rm M}- i \Phi_1^{\rm M}=  \frac{ 2 l^4s^2}{R^3} \frac{ r f''
+3 f'}{\gamma+i\omega}.\ee
In the limiting cases,  assuming $\omega \gg \gamma$ and taking
in all terms lowest  non-zero order with
respect to  $\gamma/\omega$  we get
   \be
 R_1^{\rm M}= \frac{2  l^4 s^2}{R^3}\left\{
\begin{aligned}
   &\frac{ \sqrt{\pi} k_0^{3/2}e^{-Q r/2} \cos\left(k_0 r + \frac{\pi}{4} \right)}{ \omega \sqrt{2 r}} ,~r>1/k_0
  \\
 &-\frac{\gamma}{ \omega^2}\frac{r}{(r^2+a^2)^{3/2}},~r<1/k_0
\end{aligned}
 \right.
\label{R1M}
  \ee
and
\be
 \Phi_1^{\rm M}=- \frac{2 l^4 s^2}{\omega R^3}\left\{
\begin{aligned}
   &\frac{ \sqrt{\pi} k_0^{3/2} e^{-Q r/2}\sin\left(k_0 r + \frac{\pi}{4}\right)}{\sqrt{ 2 r}},~r>1/k_0
  \\
& \frac{r}{(r^2+a^2)^{3/2}},~r<1/k_0
\end{aligned}
 \right.
\label{Phi1M}
  \ee
Finally, from Eqs.~\eqref{J12P-defenition}, \eqref{dvf}, and  \eqref{dvD}   we find  (in the lowest order with respect to $\gamma/\omega$)
\begin{align}
&R_2^{\rm M}= -\frac{l^4 s^2}{\gamma R^3} (r f'' +3 f') + c.c.
\nonumber
\\
&=\frac{2 l^4 s^2}{\gamma R^3}\left\{
\begin{aligned}
 &\frac{\sqrt{ \pi} k_0^{3/2} e^{-Q r/2}\sin\left( k_0 r+ \frac{\pi}{4} \right)}{\sqrt {2r}}, ~r > 1/k_0,
  \\
 &\frac{r}{(a^2+r^2)^{3/2} } , ~r<1/k_0 ,
 \end{aligned}
\right.
\label{R2M}
\\
 & \Phi_2^{\rm M} =0.
\label{Phi2M}
\end{align}

\section{Asymptotical values of $j_{\rm dc}$ and  $E_{\rm dc}$ for a single nanosphere}
Using Eqs.~\eqref{R1P},\eqref{Phi1P},\eqref{R2P},\eqref{Phi2P},
\eqref{R1M},\eqref{Phi1M},\eqref{R2M}, and \eqref{Phi2M}, we find asymptotical behavior  of
$j_{\rm dc}$ and $E_{\rm dc}$ with account of both plasmonic and mixed contribution
\be
j^{\rm dc} =  -\frac{\omega l^4 N_0}{ a^3}\left\{ \begin{aligned}
& A \left( \frac{r}{a} \right) +\frac{a}{R^3 k_0^2}~ C \left( \frac{r}{a}\right), ~r \ll 1/k_0 ,
\\
& \frac{\pi a^3}{r^3}
 \left[ k_0 r e^{-Qr}  + \sqrt {\frac{2}{\pi k_0 r}} \left( \frac{r}{R}\right)^3
 e^{-Q r/2} \sin\left( k_0 r+ \frac{\pi}{4} \right) \right], ~  1/k_0  \ll  r \ll \frac{\ln\left[{k_0}/
{Q}  \right]}{Q}
\\
& 6 a^3 \left( -\frac{1}{k_0^7 r^7} + \frac{1}{R^3 k_0^4 r^4} \right),  \qquad   \frac{\ln\left[{k_0}/
{Q}  \right]}{Q} \ll r,
\end{aligned} \right.
\label{jdc-full}
\ee
\be
\frac{ e E^{\rm dc}}{m} =  -\frac{\omega^2 l^4 }{ a^3}\left\{ \begin{aligned}
& \pi B \left( \frac{r}{a} \right) +\frac{a}{R^3 k_0^2}~ C \left( \frac{r}{a}\right), ~r \ll 1/k_0 ,
\\
& \frac{\pi a^3}{r^3}
 \left[ \frac{k_0 r}{2} e^{-Qr}  + \sqrt {\frac{2}{\pi k_0 r}} \left( \frac{r}{R}\right)^3
 e^{-Q r/2} \sin\left( k_0 r+ \frac{\pi}{4} \right) \right], ~  1/k_0  \ll  r \ll \frac{\ln\left[{k_0}/
{Q}  \right]}{Q}
\\
& 6 a^3 \left( \frac{5}{k_0^7 r^7} + \frac{1}{R^3 k_0^4 r^4} \right),  \qquad   \frac{\ln\left[{k_0}/
{Q}  \right]}{Q} \ll r,
\end{aligned} \right.,
\label{Edc-full}
\ee
where
 \begin{eqnarray}
  &&  A(x)=\frac{2 x}{ (1+x^2)^{3/2} (1+ \sqrt{1+x^2})}, \\
\nonumber
   &&  B(x) =\frac{2 x}{  (1+ \sqrt{1+x^2})^3}
                                 \frac{1+(3+x^2)\sqrt{1+x^2}}{(1+x^2)^2},
                                 \\
&&
C(x)=\frac{2x}{(1+x^2)^{3/2}}
\label{C}
\end{eqnarray}
\section{Expressions for   $\m J^{\rm P}_i, \m J^{\rm M}_i $  for  periodic array of nanospheres. }
\label{app:Lattice}
{For a periodic    array of nanospheres   one should replace $Z(\m r )$   with  the following sum
\begin{equation}
\sum \limits_{n,m} Z(\m r- \m r_{nm})
=\sum \limits_{n,m}\int\frac{d^2 \m q}{(2\pi)^2}Z_{\m q} e^{-i \m q  \m r_{nm}}e^{i \m q \m r },
\label{sumZ}
\end{equation}
where $Z(\m r)$ is given by Eq.~\eqref{Z} and
\begin{equation}
\m r_{nm}=d(n\mathbf{e_x}+m\mathbf{e_y})
\end{equation}
are lattice vectors of the squared array.  Next, we use the Poisson summation formula
\begin{equation}
\sum_{n,m} e^{-i\m q \m r_{nm}}=\sum_{n}e^{-i d q_x n}\sum_{m}e^{-i d q_y m}=
\left(\frac{2\pi}{d}\right)^2 \sum_{m} \delta\left(q_x-\frac{2\pi m}{d} \right)
\sum_n \delta\left(q_y-\frac{2\pi n}{d} \right).
\label{Poisson}
\end{equation}
Substituting Eq.~\eqref{Poisson} into Eq.~\eqref{sumZ} and integrating over $d^2 \m q$  we  get
\begin{equation}
\sum \limits_{n,m} Z(\m r- \m r_{nm})= \frac{1}{d^2} \sum \limits_{\m q= \m q_{nm}}
Z_{ \m q}e^{i \m q  \m r },
\label{sumZ1}
\end{equation}
where inverse lattice vectors $\m q_{nm}$ are given by Eq.~\eqref{qmn}.
}

The rectified currents $\m J_i^{\rm P} $ are given by double sums over $\m q, \m q'$
{(both $\m q$ and $  \m q'$ run over values $\m q_{nm}$),} while
$\m J_i^{\rm M} $ by ordinary ones.
For convenience of further calculations,
in  plasmonic contribution we  a   introduce
Kronecker symbol $\delta_{\m Q, \m q-\m q'}$ and   sum over $\m Q$:
\begin{align}
   & \m J_1^{ \rm P}(\m r)
   =\sum \limits_{\m Q}  e^{i \m Q \m r} \m J_{1\m Q}^{P} + c.c.
   = \frac{4 \pi^2 l^4}{d^4}\sum \limits_{\m Q}  e^{i \m Q \m r}\sum
   \limits_{\m q,\m q'} \delta_{\m Q, \m q-\m q'}
   \frac{e^{ -i(\varphi_\m q-\varphi_{ \m q'}) -a(q+q')} \omega \m q q'^2}{(q^2-k^2)(q'^2-k^{*2})}+c.c.\\
   & \m J_2^{ \rm P}(\m r)=\sum \limits_{\m Q}  e^{i \m Q \m r} \m J_{2\m Q}^{\rm P} + c.c.= \frac{4i \pi^2 l^4}{d^4 \gamma } \sum \limits_{\m Q}  e^{i \m Q \m r}\sum
   \limits_{\m q,\m q'} \delta_{\m Q, \m q-\m q'}
   \frac{e^{-i(\varphi_\m q-\varphi_{ \m q'}) -a(q+q')} \omega^2
   (\m q \m q') \m  q'}{(q^2-k^2)(q'^2-k^{*2})}  +c.c.\\
   & \m J_1^{\rm M}(\m r)=\sum \limits_{\m Q}  e^{i \m Q \m r} \m J_{1\m Q}^{\rm M}  + c.c.
   = \frac{2i \pi l^4 s^2}{d^2 R^3 } \frac{1}{\gamma+ i \omega}
   \sum \limits_{\m Q}
   \frac{e^{i\m Q \m r -i\varphi_\m Q -a Q}
    Q^2 }{(Q^2-k^2)} (\m e_x+ i\m e_y)  +c.c.\\
   & \m J_2^{\rm M}(\m r)=\sum \limits_{\m Q}  e^{i \m Q \m r} \m J_{2\m Q}^{\rm M}+ c.c.=\frac{2 \pi l^4 s^2}{d^2 R^3 \gamma  } \frac{1}{\gamma+ i \omega}
   \sum \limits_{\m Q}
   \frac{e^{i\m Q \m r } e^{ -a Q}
    \omega \m Q Q }{Q^2-k^2}   + h.c
\end{align}
 Using Eqs.~\eqref{J-lattice-expansion} and \eqref{JDCQ} we find expression for optically-induced
 dc current, which includes both plasmonic  and  mixed  contributions:
\begin{align} \label{jdc-lattice}
&\m j^{dc}= N_0\frac{4\pi^2 l^4}{d^4}  \left \{  \omega \sum \limits_{\m q,\m q' }
 \frac{\left[\m e_z \times(\m q-\m q')\right]  }{|\m q-\m q'|}
 \left(\frac{\left[\m e_z \times(\m q-\m q')\right] \m q }{|\m q-\m q'|}\right)
\frac{   q'^2e^{i(\m q- \m q') \m r} e^{ -i(\varphi_\m q-\varphi_{ \m q'}) -a(q+q')}
}{(q^2-k^2)(q'^2-k^{*2})} \right.
\nonumber
\\
 & \left.+ \frac{i s^2 d^2}{2\pi R^3}\frac{1}{\gamma+ i \omega}\sum \limits_{\m Q}
 \frac{\left[\m e_z \times\m Q \right]}{Q}
 \left(\frac{\left[\m e_z \times\m Q \right](\m e_x + i \m e_y)}{Q} \right)
       \frac{ Q^2 e^{i\m Q \m r  - i \varphi_\m Q -a Q}
        }{Q^2-k^2}
\right\}+ c.c.
\end{align}
\begin{align} \label{Edc-lattice}
   & \frac{e \m E_{dc}}{m}=\gamma  \sum \limits_{\m Q} \frac{\m Q}{Q} \left \{ \frac{ \m Q}{Q} \left[
\m J_{1 \m Q}^{ \rm P}+ \m J_{2 \m Q}^{\rm P} +\m J_{1\m Q}^{\rm M} + \m J_{2 \m Q }^{ \rm M}
   \right] \right \}  e^{i \m Q  \m r}+ c.c.
      \\
   &\approx (\text{for $\gamma \ll \omega$}) \approx \gamma  \sum \limits_{\m Q}
   \frac{\m Q}{Q} \left \{ \frac{ \m Q}{Q} \left[
 \m J_{2 \m Q}^{ \rm P} + \m J_{2\m Q}^{\rm M}
   \right] \right \}e^{i \m Q  \m r} + c.c.
   \\
   &=\frac{4\pi^2 l^4}{d^4}  \left \{ i \omega^2 \sum \limits_{\m q,\m q' }
 \frac{\m q-\m q'  }{|\m q-\m q'|}
 \left[\frac{(\m q-\m q') \m q' }{|\m q-\m q'|}\right]
\frac{   (\m q  \m q')e^{i(\m q- \m q') \m r} e^{ -i(\varphi_\m q-\varphi_{ \m q'}) -a(q+q')}
}{(q^2-k^2)(q'^2-k^{*2})} \right.
\nonumber
\\
 & \left.- i\frac{ s^2 d^2}{2\pi R^3}\sum \limits_{\m Q}
        \frac{ Q \m Q e^{i\m Q \m r   -a Q}
        }{Q^2-k^2}
\right\}+ c.c.
\end{align}
\section{Expressions for  rectified currents in the fundamental  plasmonic mode }
\label{app:fund}
For $(n,m) = (1,0),(0,1),(-1,0),(0,-1)$, we have $q=q_0=2\pi/d.$   Simple calculations yield
\begin{align}\label{j1P}
   & \m J_1^{\rm P} =  \frac{32\pi^2 l^4}{d^4}~ \frac{ \omega q_0^3
   [\sin (q_0y) \cos (q_0 x) \m e_x - \sin (q_0x) \cos (q_0 y) \m e_y]}{|q_0^2 -k^2|^2} e^{-2 q_0 a},
   \\
   \label{j2P}
   &\m J_2^{\rm P} = \frac{32\pi^2 l^4}{d^4}~ \frac{ \omega^2 q_0^3
   [\sin (q_0x) \cos (q_0 x) \m e_x + \sin (q_0y) \cos (q_0 y) \m e_y]}{\gamma|q_0^2-k^2|^2}e^{-2 q_0 a},
   \\
   \label{j1M}
   &\m J_1^{\rm M}=\frac{4\pi l^4 }{d^2 R^3}~\frac{q_0^2 \omega}{k_*^2(q_0^2- k^2)}
    [i\sin(q_0x) +  \sin(q_0 y )]
   (\m e_x + i
   \m e_y)e^{- q_0 a} +c.c.
   \\
\label{j2M}
   &\m J_2^{\rm M}= \frac{4\pi l^4 }{d^2 R^3}\frac{q_0^2 \omega^2}{\gamma k_*^2(q_0^2 -k^2)}
    [\sin(q_0x) \m e_x +  \sin(q_0 y )\m e_y] e^{- q_0 a}  +c.c.
\end{align}
Next, we substitute these equations into
Eqs.~\eqref{JDCQ} and \eqref{EDCQ}. The latter can be written in the operator form
\begin{align}
&\m j_{\rm dc}= N_0 \frac{-  \nabla ~{\rm  div} + \Delta}{\Delta}
\left( \m J_1^{\rm P}+\m J_1^{\rm M} \right),
\label{jdc-operator}
\\
\label{Edc-operator}
&\frac{e \m E_{\rm dc}}{m} =\gamma \frac{\nabla}{ \Delta} {\rm div}\left( \m J_1^{\rm P}+\m J_1^{\rm M}
+\m J_2^{\rm P}+\m J_2^{\rm M}  \right),
\\
\label{Phidc-operator}
&\frac{e \m \phi_{\rm dc}}{m} =\gamma \frac{1}{ \Delta} {\rm div}\left( \m J_1^{\rm P}+\m J_1^{\rm M}
+\m J_2^{\rm P}+\m J_2^{\rm M}  \right).
\end{align}

%
%
From Eqs.~\eqref{j1P}, \eqref{j2P}, \eqref{j1M}, \eqref{j2M}, \eqref{jdc-operator},
\eqref{Edc-operator}, and \eqref{Phidc-operator}, we find

\begin{align} \label{jdcZM-full}
&\m j _{\rm dc}  = N_0 \left \{ \frac{16 \pi^2 l^4}{ d^4}
\frac{\omega q_0^3 e^{-2 q_0 a}[\sin (q_0y) \cos (q_0 x) \m e_x - \sin (q_0x) \cos (q_0 y) \m e_y]}
{|q_0^2-k^2|^2} \right.
\\
&  \left.+ \frac{4 \pi l^4}{d^2 R^3 }~\frac{\omega q_0^2 e^{-q_0a }
[\sin (q_0 y)  \m e_x-\sin (q_0x )  \m e_y  ]}{ k_*^2 (q_0^2-k^2) } \right \} + c.c.
\nonumber
\end{align}
Close to resonance, this equation can be simplified and written in the form of Eq.~\eqref{jdc0-res} with
$\boldsymbol {\pi}$  and $\boldsymbol {\mu}$  given by Eq.~\eqref{PP} and \eqref{MM}, respectively.
We also find (for $\gamma \ll \omega$)
\begin{align} \label{EdcZM-full}
& \frac{ e\m E _{\rm dc}}{m}  =  \left \{ \frac{16 \pi^2 l^4}{ d^4}
\frac{\omega^2 q_0^3 e^{-2 q_0 a}[\sin (q_0 x) \cos (q_0 x) \m e_x + \sin (q_0 y) \cos (q_0 y) \m e_y]}
{|q_0^2-k^2|^2} \right.
\\
&  \left. + \frac{4 \pi l^4}{d^2 R^3 }~\frac{\omega^2 q_0^2 e^{-q_0 a }
[\sin (q_0 x)  \m e_x +\sin (q_0 y )  \m e_y  ]}{ k_*^2 (q_0^2-k^2) } \right \} + c.c.,
\nonumber
\end{align}
\begin{align} \label{EdcZM-full}
& \frac{ e \phi _{\rm dc}}{m}  =  \left \{ \frac{4 \pi^2 l^4}{ d^4}
\frac{\omega^2 q_0^2 e^{-2 q_0 a}[ \cos (2 q_0 x)  +  \cos (2 q_0 y) ]}
{|q_0^2-k^2|^2} \right.
\\
&  \left. + \frac{4 \pi l^4}{d^2 R^3 }~\frac{\omega^2 q_0 e^{-q_0 a }
[\cos (q_0 x)   +\cos (q_0 y )    ]}{ k_*^2 (q_0^2-k^2) } \right \} + c.c.
\nonumber
\end{align}

\end{widetext}



\begin{thebibliography}{122}
\bibitem{comment2} We assume for simplicity that embedding matrix and substrate
have the same dielectric constant $\epsilon$.
\end{thebibliography}
\end{document}